\documentclass[12pt,preprint]{aastex}

\received{2004 February 20}
\begin{document}
 
\newcommand{\kms}{$\;$km s$^{-1}$}
\newcommand{\msun}{M_{\odot}}
\newcommand{\rsun}{R_{\odot}}
\newcommand{\rat}{$R_{2}$~}
\newcommand{\dvagb}{$\Delta V^{HB}_{AGB}$}
\newcommand{\cag}{$^{12}\mbox{C}(\alpha, \gamma)^{16}\mbox{O}$~}
\newcommand{\mav}{\langle M_{HB} \rangle}
 
\title{Exploring the Upper Red Giant and Asymptotic Giant Branches: the
Globular Cluster M5}

\author{Eric L. Sandquist}
\affil{San Diego State University,Department of Astronomy,San Diego, CA 92182}
\email{erics@mintaka.sdsu.edu}

\author{Michael Bolte}
\affil{UCO/Lick Observatory, University of California, Santa Cruz, CA 95064}
\email{bolte@ucolick.org}

\begin{abstract}
We have tabulated lists of upper red giant, horizontal, and asymptotic
giant branch (RGB, HB, AGB) stars in the globular cluster M5 that
are complete to over $10\arcmin$ from the core for the RGB
and AGB samples, and $8\arcmin$ for the HB sample. The large samples
give us the most precise value of $R_2 = N_{AGB} / N_{HB}$ to date for
a single globular cluster ($0.176 \pm 0.018$). This is incompatible
with theoretical calculations using the most recent physical
inputs. The discrepancy can probably be attributed to the dependence
of observed \rat values on horizontal branch morphology.  We identify
the cluster M55 as being another possible example of this
effect. Samples of HB and AGB stars in populous clusters may provide a
means of calibrating the masses of horizontal branch stars in globular
clusters. The cumulative luminosity function of the upper red giant
branch shows an apparent deficit of observed stars near the tip of the
branch. This feature has less than a 2\% chance of being due to
statistical fluctuations.  The slope of the cumulative luminosity
function for AGB stars is consistent with the theoretically predicted
value from stellar models when measurement bias is taken into
account. We also introduce a new diagnostic $R_{clump}$ that reflects
the fraction of the AGB lifetime that a star spends in the AGB
clump. For M5, we find $R_{clump} = 0.42 \pm 0.05$, in marginal
disagreement with theoretical predictions. Finally, we note that the
blue half of M5's instability strip (where first overtone RR Lyraes
reside) is underpopulated, based on the
large numbers of fundamental mode RR Lyraes and on the nonvariable stars at
the blue end of the instability strip. This fact may 
imply that the evolutionary tracks (and particularly the colors) of
stars in the instability strip are affected by pulsations.
\end{abstract}


\section{Introduction}

Evolved-star populations in clusters have long provided important
tests of stellar evolution theory because they present us with coeval
ensembles of stars having nearly identical chemical
compositions. Because the ratio of the numbers of stars in two
different evolved phases relates to the the ratio of the intervals of
time that a star spends in the two phases, we can get a glimpse at the
stellar evolution clock. However, it is only in the most massive
clusters that we can expect to sample, and therefore test model
timescales for, stars in the shortest evolutionary
phases. Even in the most massive clusters, the huge range of stellar
densities requires that wide-field data be taken to cover the extended
halo of the cluster, while high spatial-resolution (often space-based)
photometry is gathered for the crowded cores.

This study is an analysis of evolved stars in the
massive Milky Way globular cluster M5. We have 
compiled large samples of relatively rare populations on the upper
red giant branch (RGB) and the asymptotic giant branch (AGB). Our
intent is to test whether the observed evolution conforms to our
expectations from stellar models with current physics inputs. In \S 2,
we describe the observational material we used in identifying and the
stars, and the procedures used to classify them based on their
photometric properties. In \S 3, we describe various diagnostics of
the bright populations of globular clusters, including the population
ratio \rat, the distribution of HB stars, cumulative luminosity
functions, and a new population ratio $R_{clump}$.

\section{Observational Material}\label{obs}

\begin{figure}
\plotone{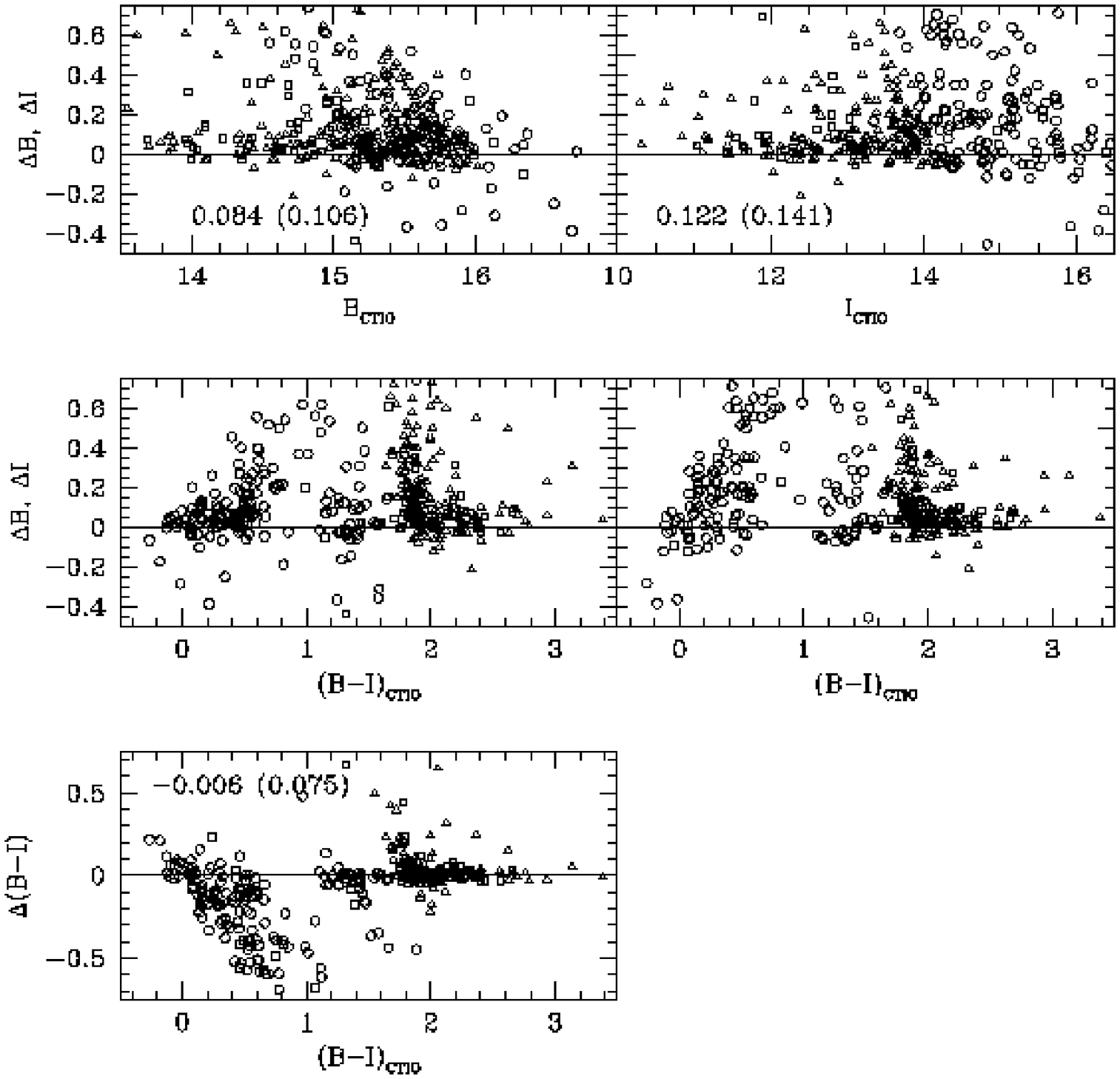}
\caption{Residuals (in the sense of CFHT minus CTIO) from the
comparison of photometry in the cluster core. Also included are the
median residual values and in parentheses the semi-interquartile range
(a measure of dispersion). Known RR Lyrae stars have been eliminated
from the plot. RGB stars are plotted with $\triangle$, AGB
stars are plotted with $\Box$, and HB stars are plotted with
$\circ$.\label{cfhtcomp}}
\end{figure}

In order to collect a nearly complete sample of bright evolved stars
in M5, we used data from a variety of sources. The primary source of
ground-based photometry was the study of Sandquist et al. (1996;
hereafter S96). That dataset was composed of $BVI$ photometry taken at
the Cerro Tololo Interamerican Observatory (CTIO) 4 m telescope and
$BI$ images of the cluster core taken using the High Resolution Camera
(HRCam) on the 3.5 m Canada-France-Hawaii Telescope (CFHT). The CFHT
data were obtained during very good seeing conditions 
so that there was little ambiguity as to the population
identifications of cluster stars. Since the 1996 study, the CFHT data
have been calibrated against the CTIO 4 m data. Fig. \ref{cfhtcomp}
presents a comparison of the photometry for the two datasets. Because
stars near the core in the CTIO sample are quite likely to be affected
by contaminating light from other nearby stars, there is a substantial
bias toward brighter magnitudes. However, the lower envelope of values
in the upper 4 plots indicates that the best measured stars have small
residuals and little or no trend with color.

\begin{figure}
\epsscale{.85}
\plotone{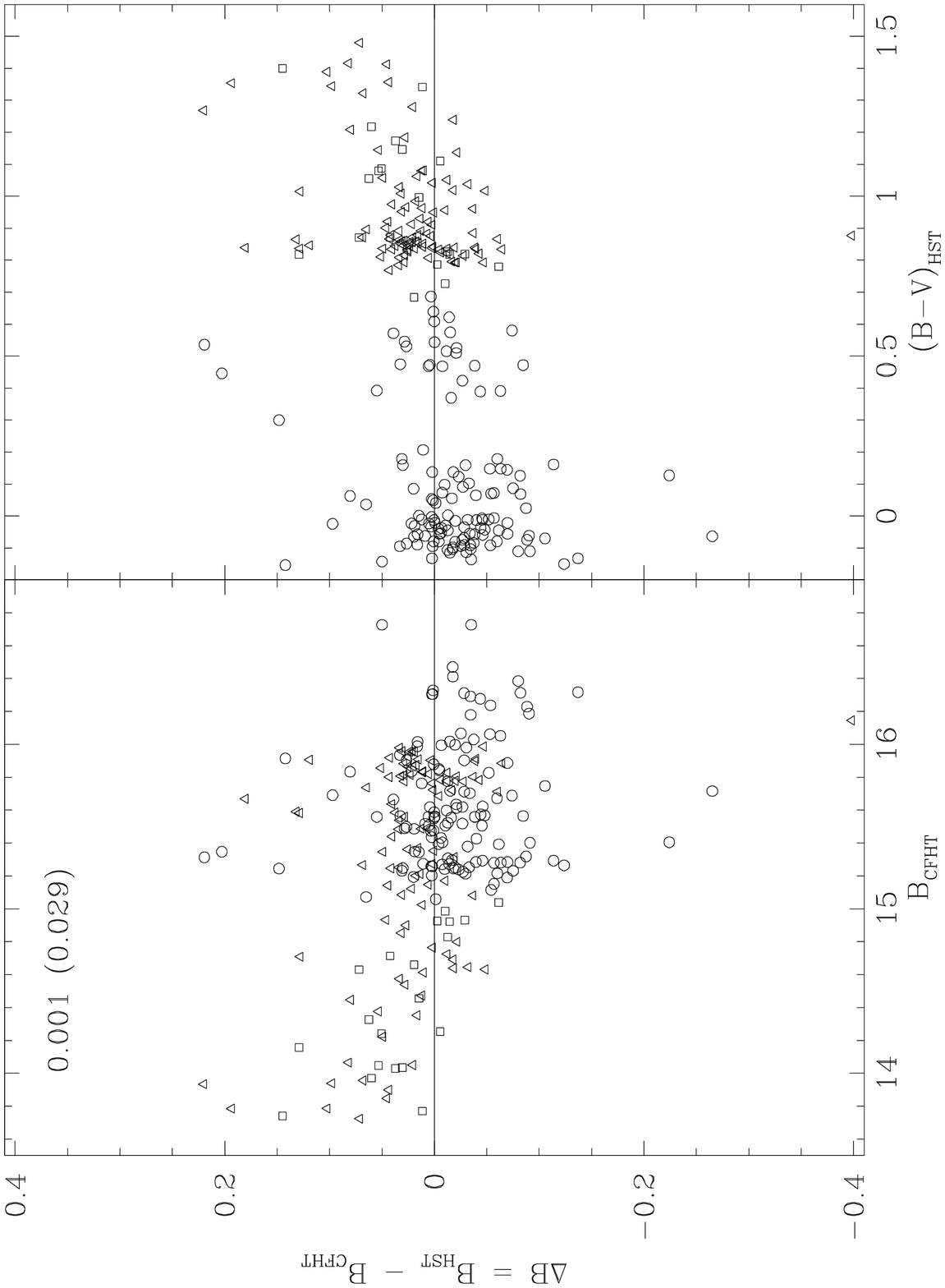}
\caption{Residuals [in the sense of HST \citep{piotto} minus
CFHT] from the comparison of photometry in the cluster core. Also
included are the median residual values and in parentheses the
semi-interquartile range (a measure of dispersion). Known RR Lyrae
stars have been eliminated from the plot.\label{hstcomp}}
\end{figure}

We have also used Hubble Space Telescope data from the cluster core
reported in \citet{piotto}. The F439W and F555W bands were
calibrated to $B$ and $V$ by that group, and we present a comparison
with our CFHT data in the core in Fig. \ref{hstcomp}. The $B$
magnitudes show evidence of either a linear trend with $B$ or a
second-order trend with color. In the course of examining the HST
dataset, we found systematic differences in the position of the RGB in
color-magnitude diagrams (CMDs) derived from different CCDs of the
WFPC2 camera. For this reason, we found that it was more reliable to
make population identifications (RGB, HB, or AGB) based on relative
position in the CMD for individual chips. Although this reduces the
number of stars on the RGB in the CMDs for each chip and makes it harder to
determine where the RGB fiducial line is, the selection of AGB stars
is generally cleaner. We have also found that the HST tabulation of
Piotto et al. apparently missed some stars whose positions fell near
where different chips overlapped. Identification numbers for the stars
in the HST dataset that are given in our final tables were composed of
the four digit ID number from the Piotto listing, with a leading digit
identifying the CCD chip used in the measurement (1 = PC, 2-4 =
WF2-4).

Positions in HST frames were derived using the IRAF\footnote{IRAF
(Image Reduction and Analysis Facility) is distributed by the National
Optical Astronomy Observatories, which are operated by the Association
of Universities for Research in Astronomy, Inc., under contract with
the National Science Foundation.} routine METRIC in the STSDAS
package, which converts pixel coordinates to sky coordinates, corrects
for geometric distortions of individual chips, and puts coordinates
from different chips on the same coordinate system. Lists from
different datasets were matched by position using the program
DAOMASTER (P. B. Stetson, private communication), which determines
six-coefficient coordinate transformations. The positions we tabulate
are given in arcseconds from the center of the cluster (taken to be
$\alpha = 15^{h}18^{m}33\fs8$, $\delta = +2\degr4\arcmin58\arcsec$,
epoch 2000.0; Harris 1996), with the x position being the $\alpha$
offset and the y position being the $\delta$ offset (positive being
higher $\alpha$ and $\delta$).

\begin{figure}
\epsscale{1}
\plotone{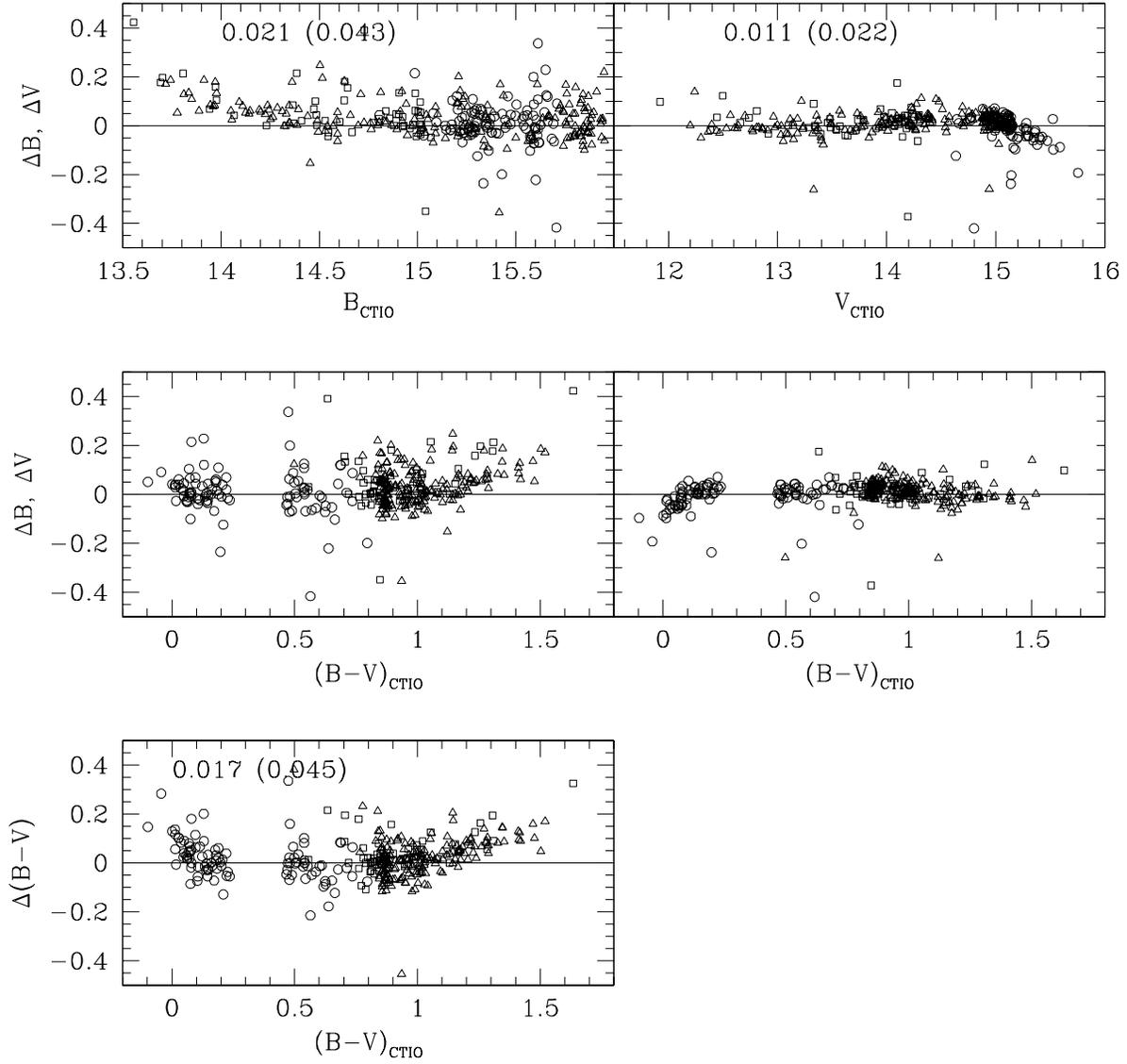}
\caption{Residuals [in the sense of Rees (1993) minus CTIO] from the
comparison of photometry in the cluster outskirts. Also included are
the median residual values and in parentheses the semi-interquartile
range (a measure of dispersion). Known RR Lyrae stars have been
eliminated from the plot.\label{pmcomp}}
\end{figure}

Proper motion information from Rees (1993) was used to eliminate field
stars from the evolved star samples. Overall this did not impact our
samples greatly because M5 is quite far from the galactic plane
($b = 46\fdg8$). The biggest impact of the proper motion information
was to verify that most stars bluer than the main line of the AGB were
field stars.  Because the Rees dataset extends well beyond the bounds
of the S96 CTIO fields, we have used his photographic data in the
outer cluster regions to identify additional stars to include in the
sample. Fig. \ref{pmcomp} shows a comparison between Rees' photometry
and our CTIO photometry. Overall, the photometry agrees well, with the
exception of slight trends with color for the brightest giants in $B$
and the faintest HB stars in $V$. For the RGB and AGB samples, the
inclusion of the Rees dataset allows us to determine complete samples
from the center of the cluster to well over $10\arcmin$. Because the
Rees data do not reach the faintest HB stars, the HB sample is only
complete from the core to approximately $8\arcmin$ from the center.

Although Figs. \ref{hstcomp} and \ref{pmcomp} make it clear that there
are systematic differences in the calibration of the photometry in
these different sets, the residuals are less than 0.1 mag in all cases
[with the exception of the tip of the RGB in $B$ in the Rees (1993)
data] and therefore this is unlikely to make any substantial changes to our
conclusions. \citet{jb} indicate that there are some calibration
issues in the S96 dataset (mean differences of $\Delta V = -0.039 \pm
0.003$ and $\Delta I = -0.021 \pm 0.003$, and position-dependent
residuals). However, because the offsets for the brightest stars are
still only at the level of a few hundredths of a magnitude, and are
considerably less serious for stars belonging to the brightest
populations (see their Fig. 6), we believe that this is not a serious
consideration for this study. Identification of the evolutionary stages of
individual stars are made relative to other stars in the same
photometric dataset where possible, and there are few cases in which
the identification is questionable. The tests of evolutionary theory
below are not strongly dependent on the brightness of
individual stars because of the samples we have
collected are relatively large.

\subsection{Population Assignment Criteria}

The assignment of a given star into the RGB, HB, or AGB groups was
based on position in the CMD. For different regions of the cluster, we
used the dataset and color-magnitude combination that provided the
cleanest separation of the evolved sequences. The most challenging tasks
were distinguishing between upper RGB and AGB stars, and identifying HB
stars at all brightness levels with almost no field or cluster star
contaminants. In all cases, RGB, HB, and AGB samples were first
selected from each individual dataset based solely on CMD position in
{\it that} dataset. The lists from different datasets were then
compared and discrepancies were reconciled, generally in favor of the
identification from the highest resolution, least noisy dataset. In
each case, we conducted positional searches near the positions of
candidates to try and ascertain the reason for the discrepancy (such
as blending and CCD flaws).

In the core of the cluster the only practical choices were
the HST or CFHT datasets. The CFHT photometry was given higher weight
than the HST photometry in regions of overlap because there was
smaller amount of scatter in the CFHT photometry on the giant
branches. High-resolution photometry of the core of the cluster is
important for selecting AGB stars because in lower-resolution
ground-based data, blends of RGB and HB stars or two RGB stars can
produce objects that appear to be AGB stars.  Although this
possibility exists in the outskirts of a cluster (meaning that we
cannot be certain that our samples are completely free of such
blends), the probability of overlap is much smaller. 

The stars in the three samples are presented in Tables \ref{agb} -
\ref{hb}. The first line for each star contains the new identification
number from this study, the offsets from the cluster center in arcsec,
and the membership probability \citep{rees} when available. Subsequent
lines present identifications and photometry from the different
catalogs: HST \citep{piotto}, CFHT (S96, calibrated for this study),
CTIO (S96, from either the $BV$ or $BI$ calibrations), and PM
\citep{rees}.  Notes on some stars are provided in the Table \ref{notes}. The
RGB star sample is truncated at the $V$ magnitude of the HB ($V =
15.15$). This falls slightly below the RGB bump.

\begin{figure}
\plotone{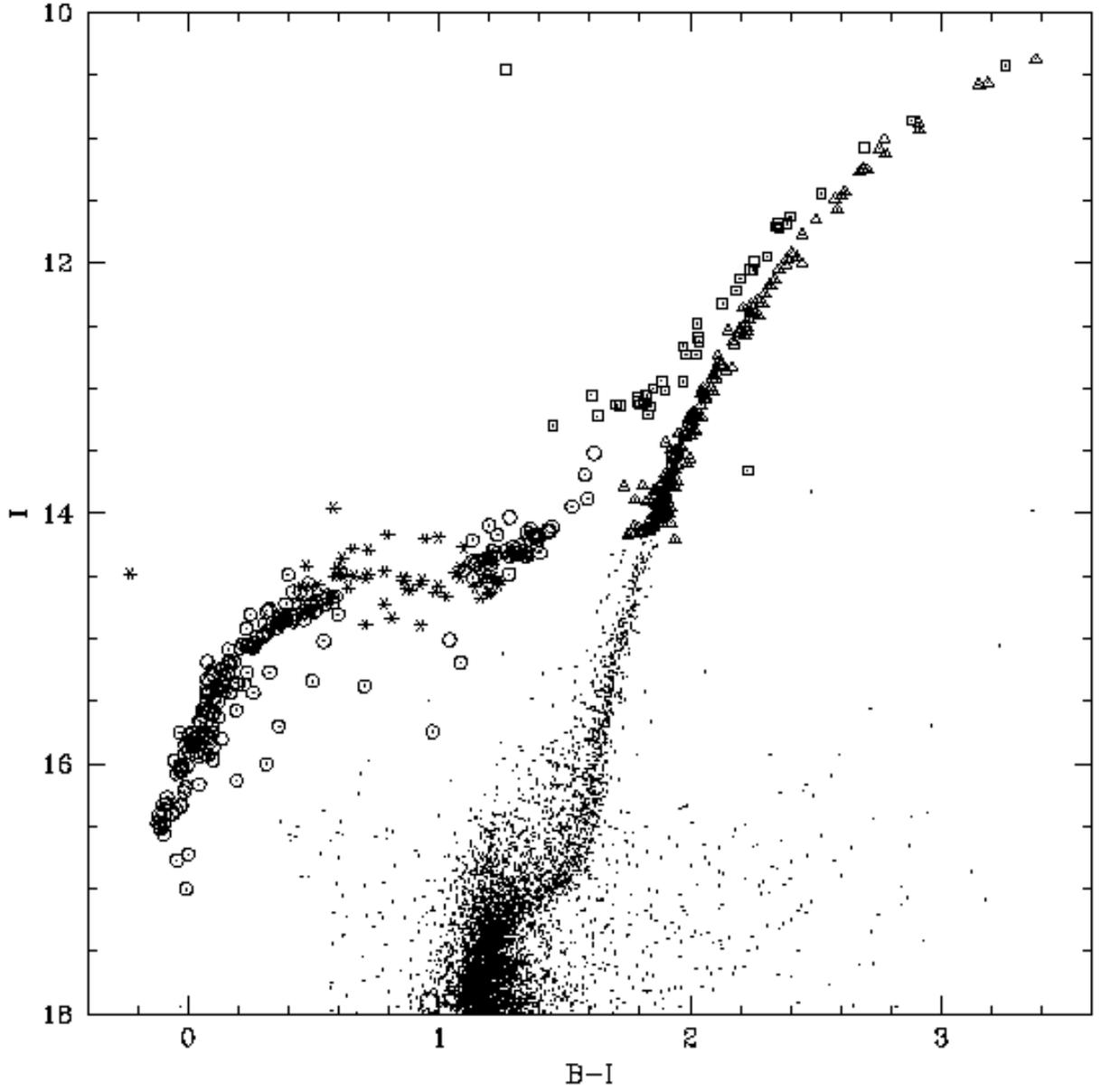}
\caption{Color-magnitude diagram for the CFHT dataset. Selected RGB
stars are plotted with $\triangle$, AGB or post-AGB stars are plotted
with $\Box$, HB stars are plotted with $\circ$, and RR Lyrae variables
are plotted with $\ast$.\label{cfht}}
\end{figure}

\begin{figure}
\plotone{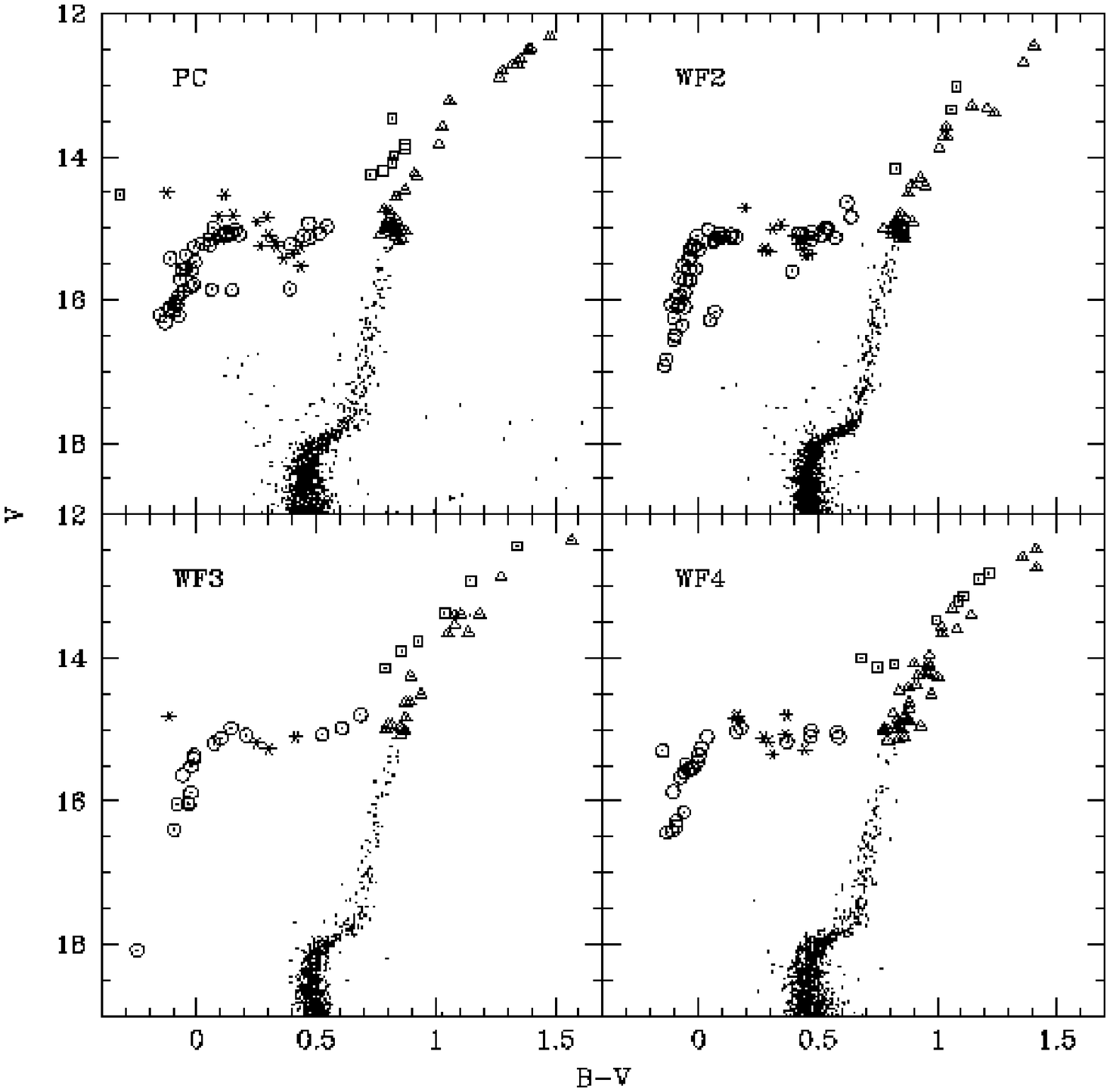}
\caption{Color-magnitude diagram for the HST dataset
\citep{piotto}. The symbols are the same as in Fig. \ref{cfht}.\label{hst}}
\end{figure}

We also must note a practical aspect to the identification of post-HB
stars.  We identify AGB stars as being those stars falling close in
color to a track nearly paralleling the RGB. Theoretical models
predict, however, that stars originating on the bluest portions of the
HB never reach the AGB \citep{dro}. These stars actually spend longer
times (up to about 100\% longer) in a double fusion shell phase than
do AGB stars that come from the reddest portions of the HB, but maintain
considerably higher surface temperatures. As a result, proper motion
information is important for the separation of these ``AGB-manqu\'{e}''
stars from foreground stars that fall in the same portion of the
CMD. As can be seen from Figs. \ref{cfht} and \ref{hst}, M5 appears to
be largely devoid of AGB-manqu\'{e} stars, which plays a role in its \rat
value, as we will see in the next section.

\begin{figure}
\plotone{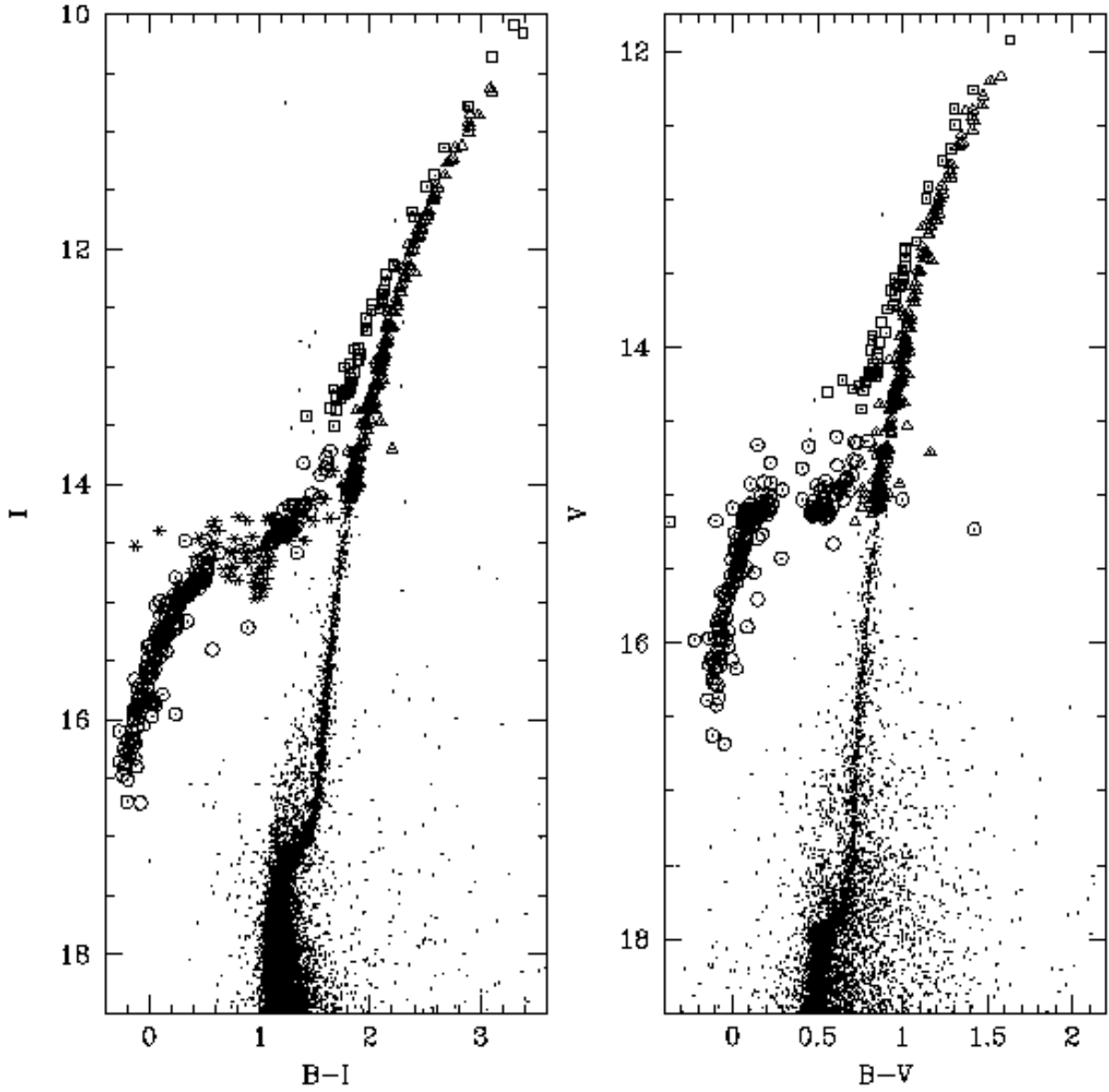}
\caption{Color-magnitude diagram for stars in the CTIO datasets that
are not in the HST or CFHT datasets. The symbols are the same as in
Fig. \ref{cfht}. Known RR Lyrae stars are not plotted for the $BV$
dataset.\label{ctio}}
\end{figure}

\begin{figure}
\plotone{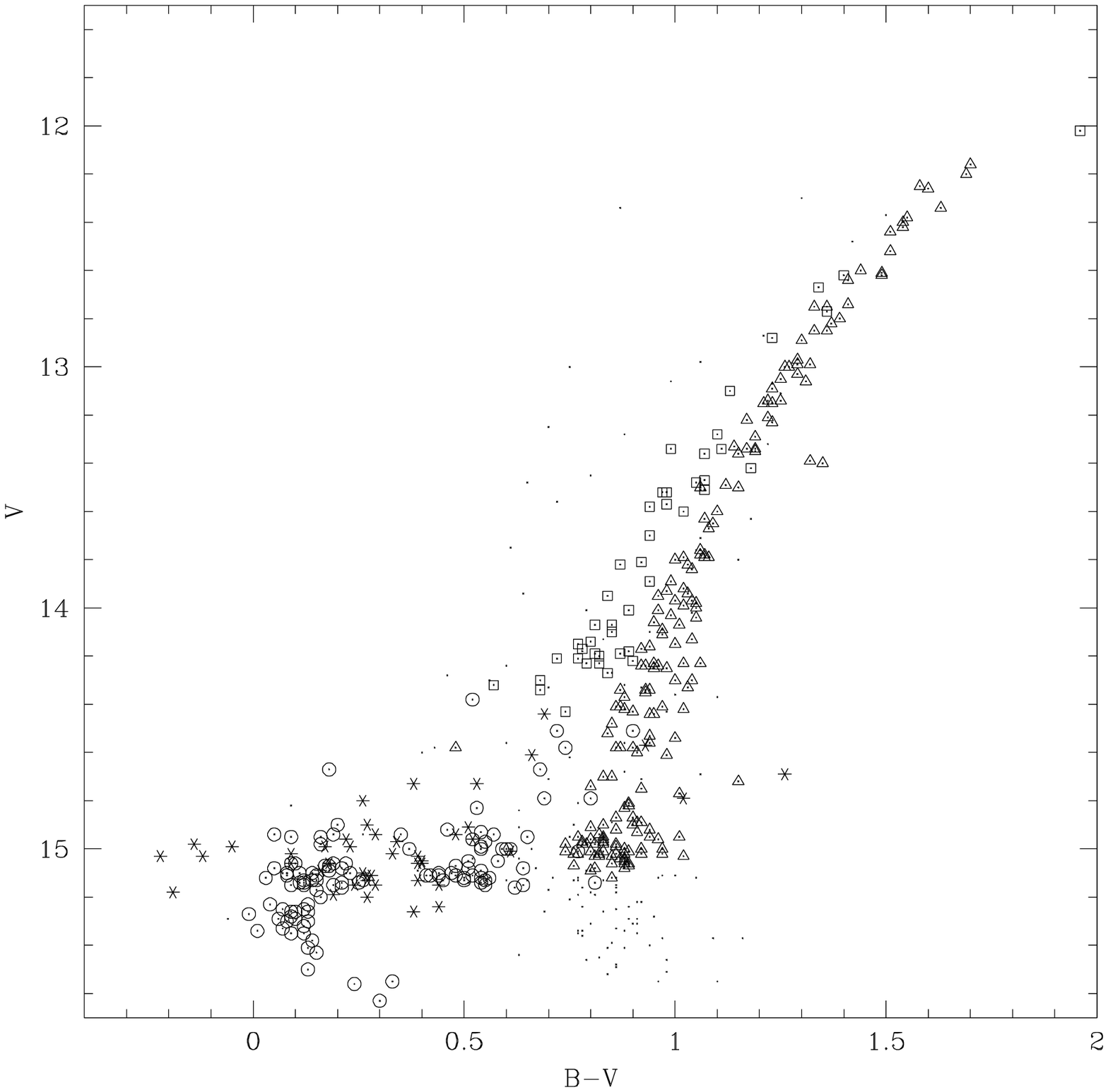}
\caption{Color-magnitude diagram for the \citet{rees} dataset. The symbols are
the same as in Fig. \ref{cfht}.\label{pm}}
\end{figure}

When photometry in multiple bandpasses was available, we examined CMDs
using different combinations of magnitude and color to help verify
identifications. Often the use of the shortest wavelength filter band
on the magnitude axis of the CMD was most helpful since this gives
both the HB and AGB/RGB a horizontal appearances in the CMD. This can
be seen in $U$, $U-V$ CMDs of the cluster (e.g. Fig. 2 of Markov,
Spassova, \& Baev 2001). CMDs of the different samples are shown in
Figs. \ref{cfht} - \ref{pm}. We put the CMDs in order of importance in
the selection. The CTIO $BI$ dataset is given precedence over the $BV$
dataset due to poorer seeing during the $V$ photometry.

We attempted to identify RR Lyrae variable stars from the literature
in order to avoid the accidental elimination of these stars from the
HB counts. The primary source of information on these stars was the
Catalog of Variable stars in Globular Clusters (hereafter, CVSGC;
described in Clement et
al. 2001)\footnote{http://www.astro.utoronto.ca/~cclement/papers.html\#catalogue}. We
have also gathered variability information on RR Lyrae variables for
the cluster from Storm, Carney, \& Beck (1991; $BV$;
intensity-weighted colors), Brocato, Castellani, \& Ripepi (1996;
$BV$; magnitude averages), Reid (1996; $VI$; magnitude-weighted
colors), and Caputo et al. (1999; $BV$; corrected colors).  To insure
that the resulting ``static'' colors were on the same scale, we
corrected the Storm et al. colors using values from Bono, Caputo, \&
Stellingwerf (1995). In the process of compiling this list, we
discovered that the variables V162 and V163 \citep{olech} were
actually the previously discovered variables V90 and V17 respectively.


\section{Diagnostics}

\subsection{The HB Distribution}

\subsubsection{HB Observed Color Distribution}

HB stars represent the sample from which the AGB stars are
born and we observationally identify the portions of the HB
that contribute the most stars to the observed AGB by examining the
distribution of HB stars. A detailed analysis of this question will
require comparison of observations with Monte Carlo simulations of HB
and AGB populations using new tabulations of theoretical models with
updated compositions and physics. For the time being, we restrict ourselves
to characterizing the HB population.

Numerous previous studies have computed HB star distributions where
the independent variable is a CMD length coordinate running from the
red end of the HB to the blue end (e.g., Crocker, Rood, \& O'Connell
1988; Fusi Pecci et al. 1993). Computer implementations of this idea
have generally used scale factors to set the relative contributions of
increments in color and in magnitude. Ideally though, we would like to
be able to relate the derived distribution function to the
distribution of HB star masses in a straightforward manner. Although
this transformation is complicated by evolution of the stars and
uncertainties in chemical composition, these uncertainties mostly
affect the average HB mass that is derived, and not the shape of the
distribution \citep{crock}. Because position on the HB is a nonlinear
function of total mass, we decided to attempt to remove this
nonlinearity using the procedure of \citet{crock}, projecting the
stars onto a theoretical ZAHB locus for an appropriate chemical
composition. This has the advantage of producing a distribution that
will have some physical meaning (something which cannot as easily be
said about $\ell_{HB}$ distributions that use fixed color and
magnitude scale factors).

\begin{figure}
\plotone{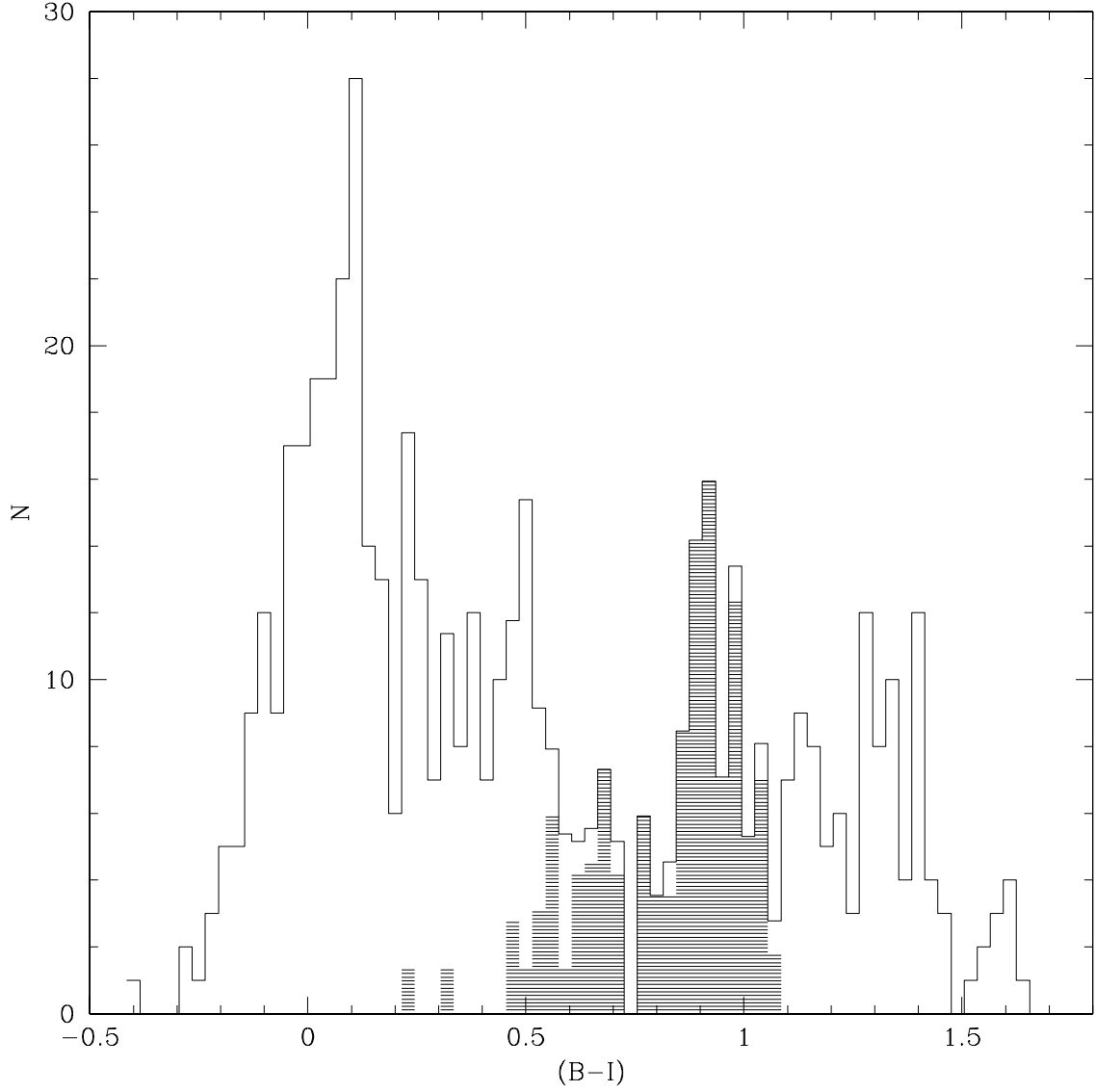}
\caption{The distribution of $(B-I)$ colors for M5 HB stars. Known RR
Lyrae stars are shown in the shaded part of the
histogram. \label{hbcol}}
\end{figure}

We elected to use our $B$ and $I$ photometry in computing mass
distributions because the $BI$ datasets cover the vast majority of the
stars, have the greatest photometric accuracy, and the $(B-I)$ color is
sensitive to temperature over the full range covered by M5 HB stars. We have
incorporated RR Lyrae variables using ``static'' colors derived from
studies listed in \S \ref{obs}. Since these studies did not use
$(B-I)$ colors, we used our CTIO datasets to do a linear interpolation
across the instability strip to convert from $(B-V)$ or $(V-I)$ colors
to $(B-I)$. For the color distribution shown in Fig. \ref{hbcol}, we
scaled the RRab and RRc portions of the histogram separately (by 1.771
and 1.38, respectively) to account for known variables (CVSGC) of the two types that do not have
measured static colors. 
For the mass distributions computed below, we
randomly chose a $(B-I)$ color from within the range covered by
variables with the same pulsation mode ($0.82 \la (B-I) \la 1.06$ for
RRab variables, and $0.46 \la (B-I) \la 0.8$ for RRc variables) if the
variable did not have a tabulated static color. The pulsation mode was
taken from the CVSGC tabulation. Because the RR Lyrae instability
strip covers a relatively small range of mass in the theoretical
models ($\sim 0.01 \msun$), the exact method used is not critical to
the overall shape of the distribution.

\subsubsection{HB Mass Distribution}

\begin{figure}
\plotone{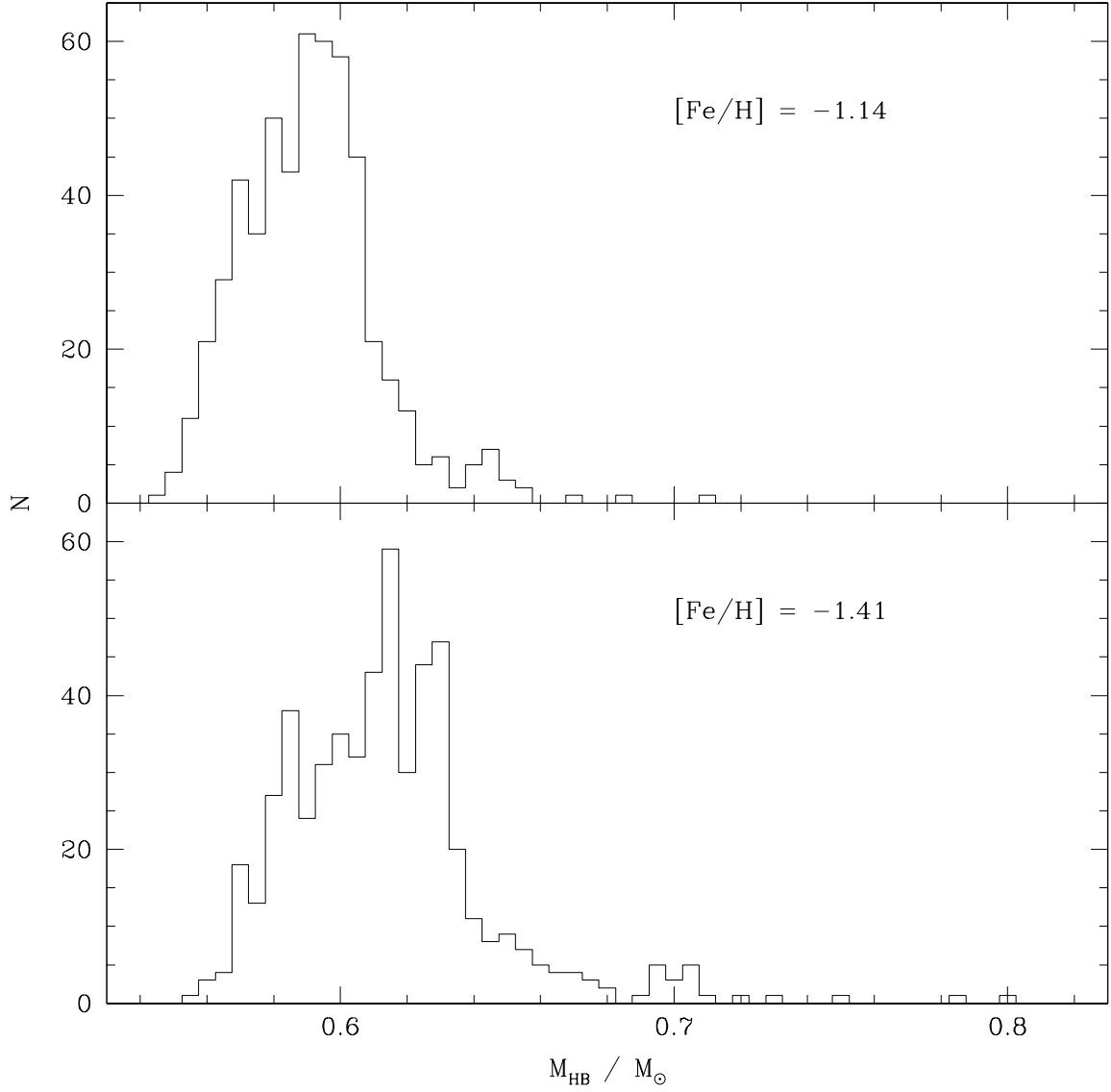}
\caption{HB mass distributions computed using $\alpha$-element enhanced
models of \citet{vdb}.\label{hbm}}
\end{figure}

We plot computed HB mass distributions in Fig. \ref{hbm} using ZAHB
models from \citet{vdb} for enhanced $\alpha$-element abundances using
two [Fe/H] values roughly corresponding to tabulated values from
\citet{zw} and \citet{cg} and a reddening E$(B-I) = 0.086$ (Schlegel,
Finkbeiner, \& Davis 1998). We note that with this reddening the blue
tail of the HB observations are not a good color match to the models.
Partly the discrepancy
seems to result from calibration errors for the CFHT data at the blue
end of the HB. This should not be an issue for the mass distributions
because the distributions are derived from {\it projections} to the
ZAHB.  For [Fe/H] $= -1.41$, we find $\mav = 0.614$ with $\sigma_{M} =
0.031$, while for [Fe/H] $= -1.14$, we find $\mav = 0.591$ with
$\sigma_{M} = 0.021$. The $\sigma_{M}$ values are somewhat misleading
though because the distributions are asymmetric.

The identification of the portion of the HB containing the peak of the
mass distribution and the majority of stars is a robust result of this
procedure, independent of the exact values of the star masses. The
peak of the distribution falls at $(B-I)_{peak} = 0.50$ (right at the
blue edge of the instability strip), while the median color value is
$(B-I)_{med} = 0.43$. We can characterize the width of the HB color
distribution in several ways as well. From looking at the HB mass
distribution, and identifying by eye where the distribution drops off
most quickly, we can crudely bracket the range of colors containing
the majority of stars. In this way, we find that the red edge of the
mass distribution at $(B-I) = 1.05$ (essentially the red end of the
instability strip), and the blue edge falls on the blue tail at $I
\approx 16.2$ (although this is more poorly defined). The stars on the
red side of the instability strip form a tail in the mass distribution
largely because HB color changes much more slowly with mass there.
Alternately we can identify the middle 68.3\% of the HB sample
(corresponding to $1 \sigma$ points of a Gaussian distribution).  In
this way we find a red bound at $(B-I) = 1.14$ in the blue half of the
constant red HB, and a blue bound at $I \approx 15.8$ on the blue tail.
These two methods give similar results.

\subsubsection{Oosterhoff Group Considerations}

The color distribution emphasizes a peculiar property of M5: a
majority of M5's HB stars fall on the nonvariable blue HB, but M5 is
an Oosterhoff group I cluster with the expected excess of RRab type
variables ($N_c / N_{ab} = 39 / 91 = 0.43$; CVSGC), which reside on
the red side of the instability strip. In fact, M5 has one of the
bluest HBs of the Oosterhoff group I clusters in the Galaxy. Recently
\citet{jurc} presented evidence that the dominant Oo I population in
M3 is composed of stars in the early stages of post-ZAHB evolution,
and that stars that have begun to evolve redward appear to have an Oo
II-like population. Jurcsik et al. reference the hysteresis hypothesis
of \citet{van} that proposed that switches between fundamental and
overtone pulsation modes occur at different temperatures depending on
the direction of the star's evolution in color. This may also be the
case for M5: its Oo I variable population derives from the fact that
the ZAHB in the instability strip {\it is} populated as a result of
the cluster's large dispersion in $M_{HB}$. However, this does not
explain why the distribution of static colors for the variables is
biased toward the red half of the instability strip. M5 is one of the
only Oo I clusters for which the number of stars per color interval
should be rising as you go from the red end of the instability strip
to the blue end, but the cluster still has many fewer RRc variables
than RRab variables. Theoretical evolution tracks do not diverge from
this portion of the HB toward the red and blue. This seems to imply
that HB stars are being kept out of the range of colors that are
normally occupied by RRc variables. In other words, the presence of
pulsation seems to affect the evolutionary track of the star.

Our discussion has focussed on published values of static colors; however,
there is no reason to believe that this affects the conclusion. The
larger number of fundamental mode pulsators is undeniable, and the
vast majority of those RR Lyrae stars having static colors fall in two
distinct color ranges sorted by the pulsation mode. The fraction of
variables pulsating in the first overtone in other globular clusters
also shows very little dependence on HB morphology (e.g. Castellani,
Caputo, \& Castellani 2003).

\subsubsection{Unusual HB Stars}

In order to facilitate the identification of unusual stars, we plot a
CMD for the combination of the CFHT and CTIO datasets in
Fig. \ref{bothbi}. This is the largest sample with a common
photometric calibration (although a slight mismatch in the colors of
the bluest HB stars in the two datasets is evident).

\begin{figure}
\plotone{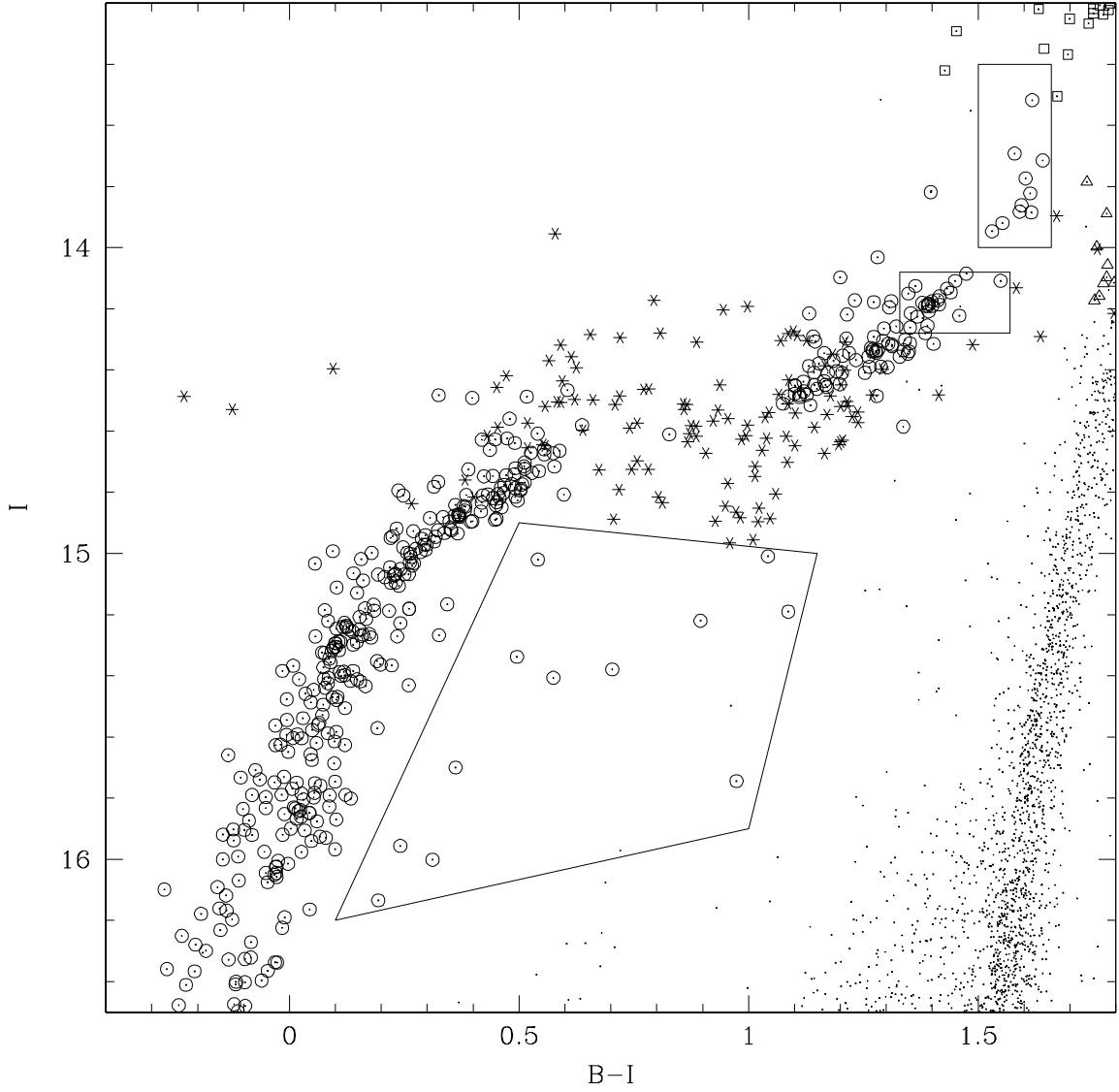}
\caption{Zoomed CMD for HB stars identifying two groups of extremely
red HB stars (ER HB) and stars likely to be unresolved blends of blue
HB stars and faint RGB stars. \label{hbzoom}}
\end{figure}

In examining HB stars, we have found several groups of stars with
unusual magnitudes and colors even in the best datasets.  There are
two separate groupings of stars redder and brighter than the most
populated portion of the red HB in M5. These two groups may correspond
to groups in M3 that were given the classification ``ER'' by
\citet{ferrm3}. The brighter of the two groups ($I < 14$) contains 11
stars (IDs 13, 28, 30, 53, 110, 118, 137, 141, 168, 239, 392) and is
at the extreme red end of the HB, noticeably disconnected from the
rest of the HB. The fainter of the two groups ($14.08 < I < 14.28$;
see Fig. \ref{hbzoom}) is less noticeably separated from the red HB
and contains 14 stars (IDs 34, 42, 62, 63, 67, 76, 169, 198, 360, 389,
412, 425, 434, 444). The two groups contain about 4.5\% of the total
HB sample.

\begin{figure}
\plotone{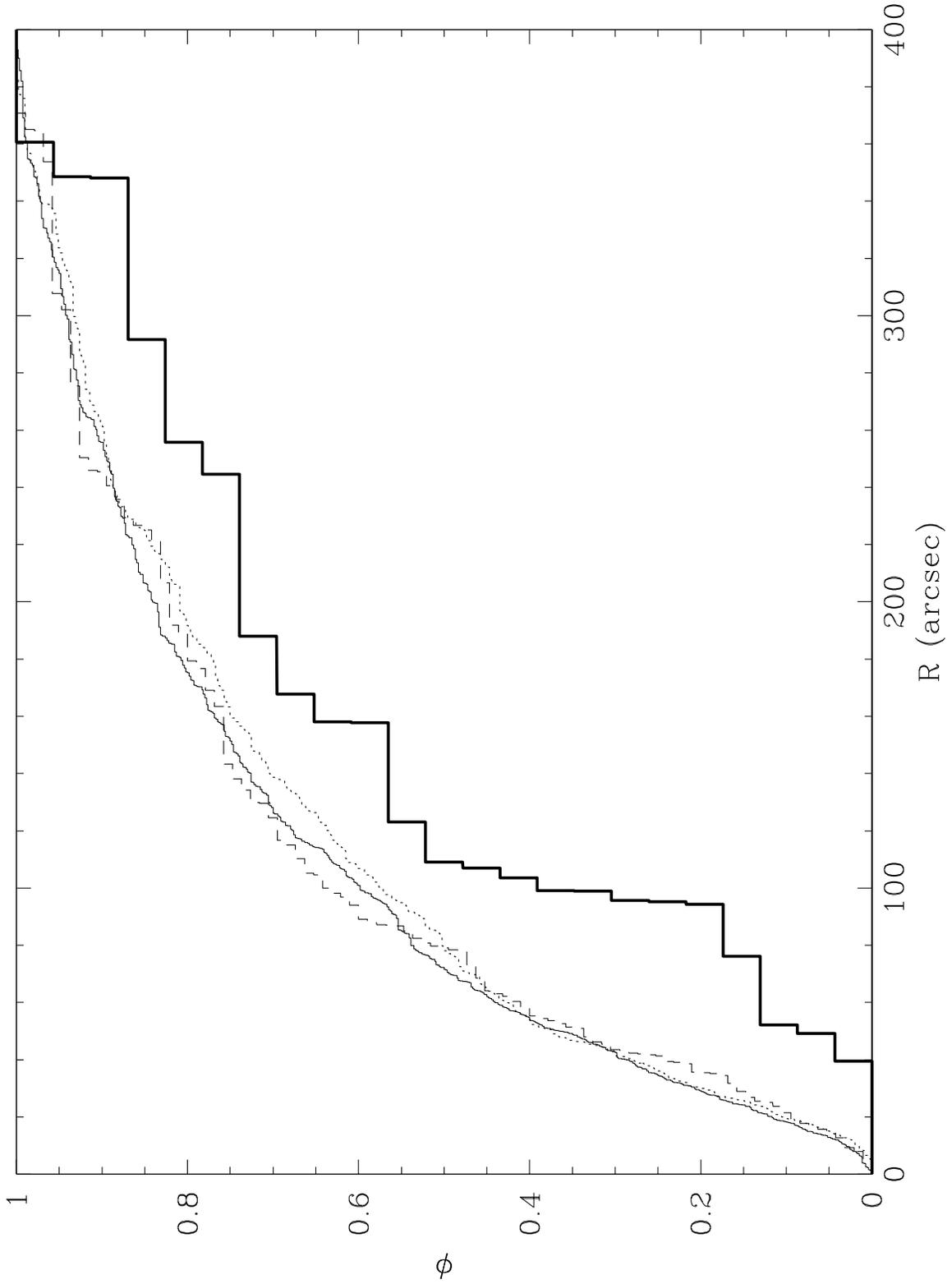}
\caption{Cumulative radial distributions of RGB ({\it dotted line}),
HB ({\it solid line}), AGB ({\it dashed line}), and extreme red
HB ({\it thick solid line}) stars.\label{crds}}
\end{figure}
\begin{figure}
\plotone{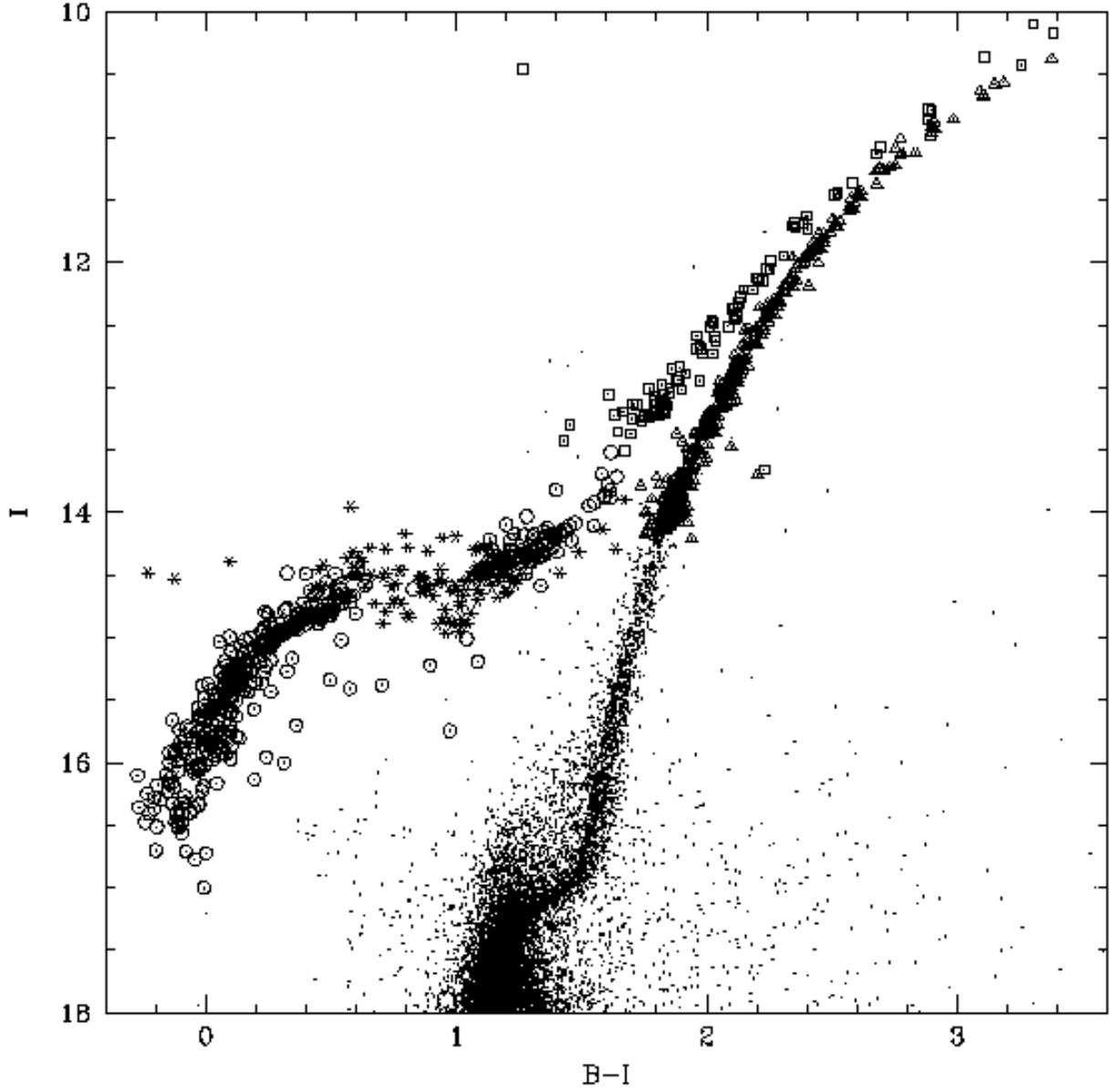}
\caption{Combined color-magnitude diagram for stars in the CFHT and
CTIO datasets. Stars measured in both datasets are plotted with
photometric values from the CFHT dataset. The symbols are the same as
in Fig. \ref{cfht}. \label{bothbi}}
\end{figure}

\citet{fusi92} hypothesized that the ER HB stars are the progeny of
blue straggler stars --- because blue stragglers are believed to have
mass larger than turnoff mass stars, when they evolve to the core He
fusion phase, they would tend to have higher masses and redder colors
than the majority of HB stars. \citet{ferrhst} provide some evidence
that the radial distribution of ER stars in M3 parallels that of the
stragglers. \cite{ferrm80} find than the cumulative radial
distribution of the ER HB stars parallels that of the blue stragglers,
and both are more centrally concentrated than the combined RGB and HB
sample. For M5, we find that the cumulative radial distribution of the
two groups of ER HB stars is actually {\it less} concentrated than the
HB stars (Fig. \ref{crds}). From a Kolmogorov-Smirnov test, there is
only a probability of 0.0014 that the two samples come from the same
distribution. (The redder of the two groups of ER HB stars is too
small to make a statistically significant comparison.) However, 10 of
these stars are concentrated of these stars within about $1\arcmin$ of
the edge of the high resolution HST and CFHT fields, so that there is
a good chance some of these are unresolved blends of an HB and an RGB
star.  While blends are unlikely to explain other identified ER HB
stars (particularly when the ones forming a tight group in the CMD),
higher resolution photometry of the near-core region is needed.

We have also identified 13 stars fainter than the horizontal portion
of the HB and redder than the blue tail (see Fig. \ref{hbzoom}). All
of these stars are found within $100\arcsec$ of the core. As with the
ER stars, several (IDs 81, 95, 302, 317) are found just outside the
HST and CFHT fields, and so may be blends of a blue HB star and a star
on the faint half of the RGB. Others (IDs 177, 243, 316) are found
near the cluster core and were only measured in the CFHT field, so
that they may also be blends. However, others (IDs 107, 201, 221, 250,
268, 276) were measured in both the HST and CFHT fields, and deviate
in similar ways in both datasets. Blends or unresolved binaries
involving a blue HB star and a star on the lower RGB are one
possibility.  In a few of the cases, it is also possible that the
stars are extremely bright blue stragglers.  A less likely possibility
for a subset of these stars (IDs 268 and 317) is that there may be
unidentified RR Lyrae stars caught in a particular part of the
pulsation cycle.

Two stars are found in the RR Lyrae instability strip in different
datsets. HB star 340 falls within the instability strip in the HST
dataset, and within the area of the CMD covered by RR Lyrae stars
measured in the CFHT datset. As a result, we flag it as possible RR
Lyrae star. Star 132 is relatively near the core, but falls within the
instability strip in the CTIO dataset. It may be a blend of an blue HB
star and and giant star.


Finally we note that there is one star in the HST and CFHT datasets
(ID 341) several magnitudes below the blue end of the HB that may be
an extreme blue HB star. In spite of the fact that it is observed in
the core of the cluster, it may still be a field star, or a cluster
member in an evolutionary phase following the AGB. The star is
substantially bluer than the main sequence in both the $B-V$ and $B-I$
colors. We cannot distinguish between these possibilities with the
datasets discussed here.


\subsection{The Population Ratio $R_{2}$}

\subsubsection{Theory}

The population ratio $R_{2} = N_{AGB} / N_{HB}$ 
is a diagnostic of the relative durations of the
helium fusion phases of stellar evolution (Buzzoni et al. 1983). As
discussed by several groups (e.g. Straniero et al. 2003; Renzini \&
Fusi Pecci 1988), this ratio is most affected by the physical
processes occurring in the cores of stars during the horizontal branch
(HB) phase. The largest uncertainties in theoretical predictions for
this ratio are the rate for the \cag
reaction and details of the mixing processes occurring at the outer
boundary of the convective core. Even with these uncertainties,
observed values of \rat are sufficient to rule out the possibility
of the so-called ``breathing pulses'' seen in early models of AGB
stars. Breathing pulses mix fresh helium into the core and hence
prolong the HB phase, resulting in low \rat values. As a result of
\rat measurements, ad hoc algorithms have been introduced into the
models to suppress these pulses (e.g. Cassisi et al. 2001).

There are observational influences on the \rat ratio that have been
mostly ignored up to this point. Typically, a HB star becomes brighter
and redder as it evolves toward the AGB (although the star may make
smaller excursions blueward and redward in color before leaving the HB
entirely). Theoretical models (e.g. Dorman, Rood, \& O'Connell 1993)
showed that stars evolving off of the bluest extensions of a HB may
turn blueward before they reach the portions of the CMD most heavily
populated by AGB stars. Generally only stars less massive than about
$0.53 \msun$ will completely avoid the AGB, although the exact value
depends on chemical composition. Slightly redder HB stars reach the
upper AGB, but turn off toward the white dwarf cooling sequence before
reaching the tip of the AGB. HB stars that are slightly redder still
start the AGB phase in the AGB clump, which results in a
large increase in AGB lifetime. Clearly the HB morphology of a
globular cluster will affect the number of AGB stars that are
present. 

\begin{figure}
\plotone{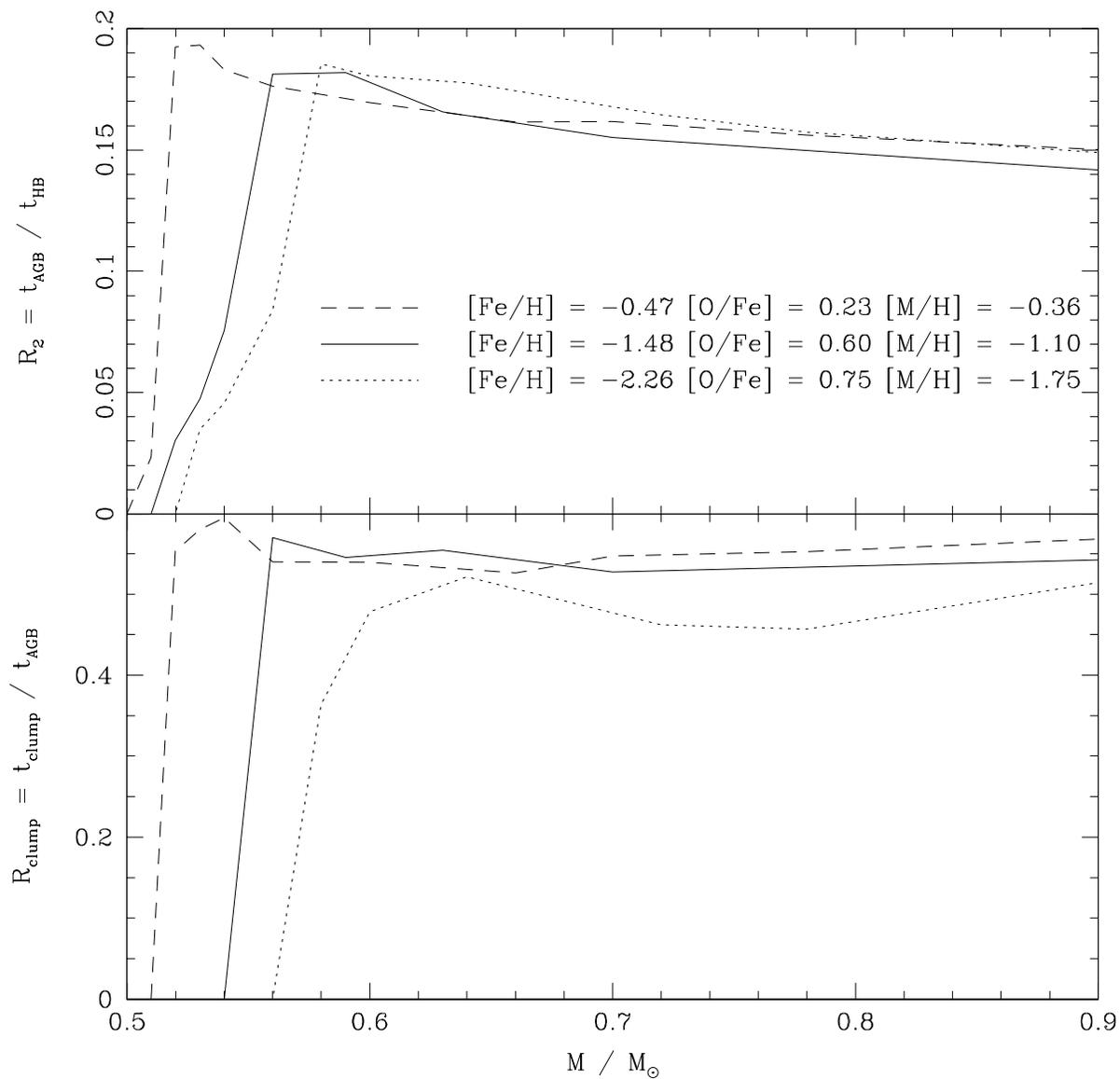}
\caption{Timescale ratios involving AGB phases from the
oxygen-enhanced models of \citet{dro}. {\it Upper panel:} ratio of
durations of the AGB and HB as a function of star mass. {\it Bottom
panel:} ratio of the durations of the AGB clump phase and the total
AGB phase.\label{rvals}}
\end{figure}

An important point of this discussion is that ``second parameter''
effects on the HB morphology can cause differences in the \rat ratio
for globular clusters of the same metallicity. As illustrations, we
present sample single-star \rat values calculated from the models of
Dorman, Rood, \& O'Connell (1993) in Fig. \ref{rvals}. We should keep
in mind that the Dorman et al.  models have physical inputs that are
somewhat out of date: oxygen-enhanced compositions (rather than
$\alpha$-element enhanced), a somewhat low initial $M_{core}$
[compared to Catelan et al. (1998), for example], and potentially an
initial envelope helium abundance $Y_{env}$ that is not consistent
with other observational constraints \citep{swei}. However, the
general features of this figure should not change. This leads to at
least two interesting questions. First, is it possible to observe
differences between the AGB populations of clusters whose HB stars
show second parameter effects?  Second, is it possible to predict a
cluster's \rat value given its metallicity and particular HB
morphology, or does the second parameter affect the AGB as well?

\subsubsection{Observations}

From our dataset, we find $N_{AGB} = 98$ and $N_{HB} = 557$ for the
total sample, excluding stars found only in the \citet{rees} dataset.
Thus, we have $R_{2} = 0.176 \pm 0.018$, where the error is computed using
\[ \sigma^{2}(R_2) = R_2 / N_{HB} .\] 
This error formula was derived 
under the assumption that stars that have passed the He flash can only
have two identifications (HB or AGB), and therefore are described by
the binomial distribution with the probability of being in the AGB
phase $p = N_{AGB} / (N_{HB} + N_{AGB}) = R_2 / (1 + R_2)$. The resulting error
$\sigma^2(R_2)$ is smaller than what is derived from Poisson errors by a factor of $(1+R_2)^{-1}$. The large
samples of stars have allowed us to substantially reduce the error on
the determination.

\begin{figure}
\plotone{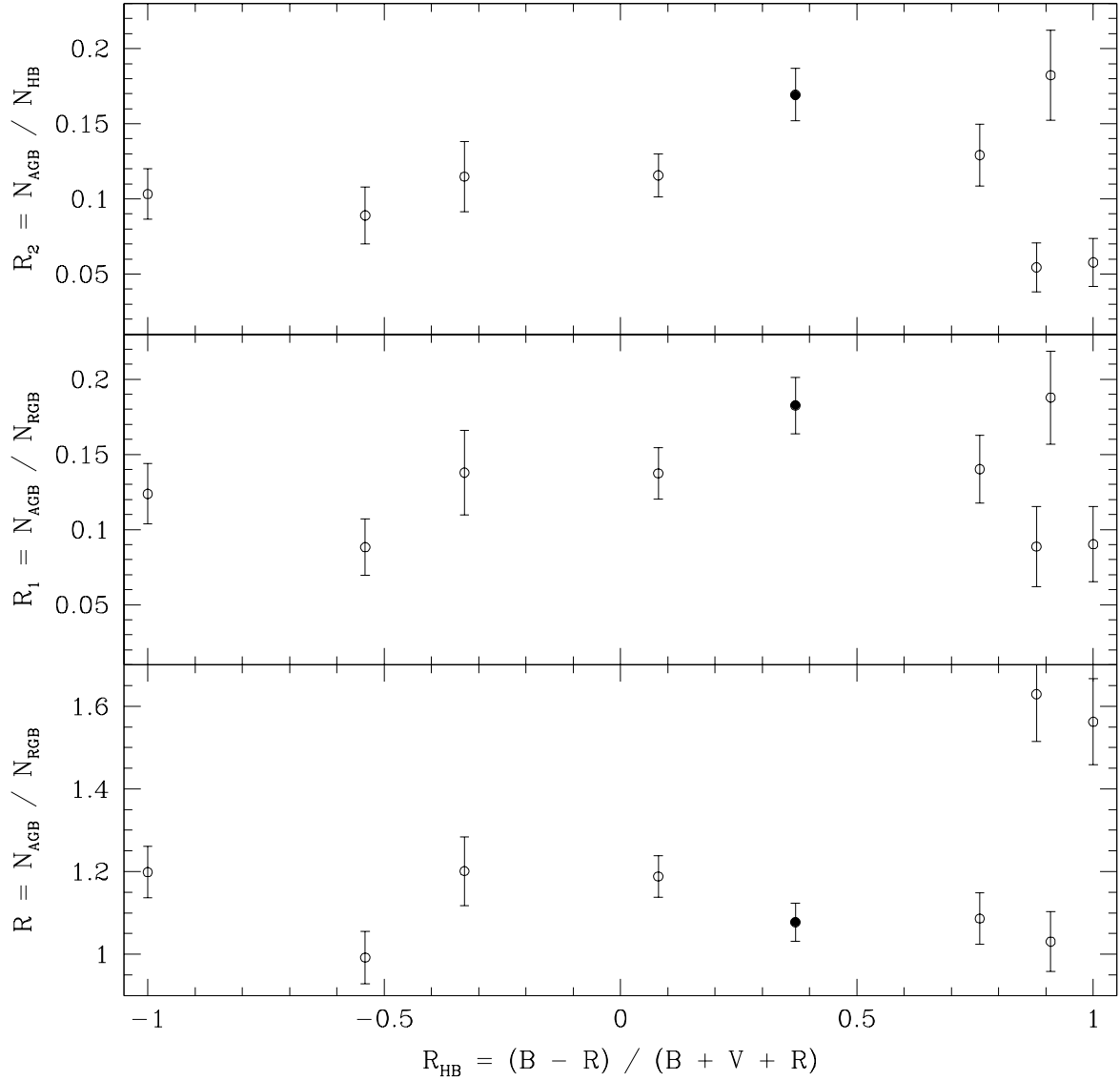}
\caption{Population ratios for globular clusters having more than 200
HB stars in the tabulation of \citet{sand}. $R_{HB}$ values come from
the tabulation of \citet{ldz}. From left the right the clusters are:
47 Tuc, NGC 2808, NGC 1851, M3, M5, M53, M30, M55, and NGC
6752.\label{robs}}
\end{figure}

The maximum single-star \rat values from \citet{dro} do agree with the
observed value for M5. However, more recent models by \citet{cass03}
with updated \cag rate and equation of state produce a value $R_{2} =
0.12$, significantly lower than the \citet{dro} and \citet{cass}
models.  The newer theoretical results are out of agreement with the
\rat value for M5 by more than $3 \sigma$, although they are in better
agreement with values from the majority of clusters with reasonably
large measured samples. In the tabulation of \citet{sand}, there were
8 clusters with more than 200 measured HB stars (see
Fig. \ref{robs}). Excluding M5, we compute $\langle R_2 \rangle =
0.106 \pm 0.038$. For the four clusters with redder HB morphologies
than M5, $\langle R_2 \rangle = 0.106 \pm 0.011$.  So, while the most
recent theoretical models are in agreement with values from the
globular clusters with the largest tabulated samples, M5 has a value
that appears to be out of agreement with other clusters and with the
theoretical values. In fact, the \rat value is one of the highest seen for 
a globular cluster to date. That M5 has a high \rat value has been
apparent even in smaller subsamples (see Table 7 of S96).

The effect that HB morphology has on numbers of AGB stars can be crudely
seen using the ratio $R_{HB} = (B - R) / (B + V + R)$ (Lee, Demarque, \&
Zinn 1994), where $B$, $V$, and $R$ are numbers of blue, variable, and
red HB stars. M5 has $R_{HB} = 0.37$ (this paper), while the slightly
more metal-poor cluster M3 has $R_{HB} = 0.19$. 
A preliminary determination of \rat for M3 gives a value of
$0.136\pm0.016$. Overall M3 has a redder HB morphology than the more
metal-rich M5, and has an \rat value that is more consistent with
typical theoretical predictions. However, the differences in \rat
values are not necessarily as simple as a shift in the peak color of
the HB distribution because the color distributions of the two
clusters overlap to a large extent. In addition, the competing effects
of HB morphology and metallicity mean that the masses of HB stars in
these two clusters are likely to be similar. (Higher metallicity makes
the ZAHB mass at a given color smaller, while a bluer HB morphology at
a given metallicity generally means a lower mean HB mass.) Catelan
(2000) computed synthetic HBs for M5, and found that a mean mass
$\langle M_{HB} \rangle = 0.633 \msun$ and a mass dispersion
$\sigma_M = 0.025 \msun$. In a similar study for M3, \citet{cat02}
finds $\langle M_{HB} \rangle \approx 0.64 \msun$ and $\sigma_M \approx
0.02 \msun$ (although they state that there appears to be a systematic
difference in $\langle M_{HB} \rangle$ as a function of cluster
radius). So although the mean mass and mass dispersion for M3 and M5
appear to be similar, M5 has a bluer HB and more AGB stars.

\subsubsection{N$_{\rm AGB}$/N$_{\rm RGB}$}

Theoretical models predict that the bluest HB stars that reach the AGB
spend a somewhat longer time in the AGB phase \citep{cass} than redder
HB stars due to smaller envelope mass and weaker fusion shell
sources. [This last point was mentioned briefly by \citet{dr}.]
Models indicate that the stars with maximal AGB lifetimes originate on
the blue tail of the HB, where the HB luminosity has dropped by about
a factor of 2 from what it is in the instability strip. As a result,
there should also be an enhancement of the numbers of AGB stars
relative to numbers of stars in other phases of evolution. Using the
numbers of RGB stars tabulated in \citet{sand}, we examine in
Fig. \ref{robs} the population ratio $R_1 = N_{AGB} / N_{RGB}$, where
$N_{RGB}$ is the number of RGB stars more luminous than the HB. The
relative positions of the clusters in a plot of $R_1$ versus $R_{HB}$
do not change --- M5 still has a higher value than clusters with
redder morphologies, which supports the idea that M5 stars spend
longer times in the AGB phase than those in clusters with redder HB
morphologies. This indicates that M5 may be anomalous in some way,
which would mean that input physics for the stellar models is {\it not} the
reason for the difference.

We note that the two clusters with the most extended blue HBs (NGC
2808 and NGC 6752) have \rat $< 0.1$, consistent with the idea that a
significant fraction of the stars are avoiding the AGB phase. The
cluster M30 also has a low \rat value, but a much more compact blue
HB. M30's metallicity is low, however, which makes it more likely that
some of its bluest HB stars would also miss the AGB (see
Fig. \ref{rvals}). The cluster M55 stands out as having high $R_1$ and
\rat values that are similar to those found for M5. Overall, M55 has a
compact blue HB with the HB distribution peaking near the bright end
of the blue tail. It also has a considerably lower metallicity than
M5, which probably means that the change in HB track morphology will occur at
higher temperatures as well. This should be checked with a study using
synthetic HB and AGB populations.

Unfortunately, there are no sets of HB and AGB star models tabulated
in the literature for the most recent physical inputs.  As a result we
are unable to gauge the exact effect of the HB morphology of M5 on the
\rat value. If the \citet{dro} models are any guide, then the \rat
value would be increased if the majority of the HB stars have masses
giving them proportionately long AGB phases.  The large \rat value
found for M5 indicates that only a small fraction of the observed HB
stars are likely to avoid having an AGB phase, which makes it possible
to place a color constraint on where the morphology of the post-HB
evolutionary tracks change. Stars that do not reach the AGB still
spend a large amount of time brighter than the HB and bluer than the
AGB. Because we have found {\it no} cluster members significantly to
the blue of the heavily-populated AGB, the post-HB evolutionary tracks
are not likely to change until $I \ga 16.4$, at the observed tip of
M5's HB. In addition, because the time spent on the AGB changes
rapidly with decreasing mass, it might be possible in the future to
calibrate HB star masses using this feature. It may also be possible
to empirically determine which HB stars have the maximum relative AGB
lifetime $t_{AGB} / t_{HB}$.


The examination of large AGB star samples gives us a way of testing
the morphology of HB tracks to see whether stars from the most
populated portions of the HB really do reach the AGB and whether they
spend the predicted amount of time there (relative to the time spent
on the HB).  By examining clusters with bluer HB morphologies or lower
metallicity, we can attempt to empirically identify the point on the
HB at which the stars no longer have an AGB phase. Although there
should be smaller differences in \rat values between clusters with red
HBs and those with intermediate morphologies, it may also be possible
to identify those stars which have AGB phases of maximum duration.


\subsection{The Luminosity Functions and Evolutionary Timescales}\label{clf}

The luminosity function (LF) of red giant branch stars is sometimes
used to test theoretical predictions for evolution timescales of
evolved stars. Although with AGB stars we are dealing with smaller
samples than are available for RGB stars, we can extract evolutionary
information using the cumulative LF. It is well-known that the
logarithm of the cumulative LF for stars on the upper RGB is linear
with magnitude \citep{fusi}. The structure of AGB stars is similar to
that of RGB stars with the hydrogen fusion shell providing the
majority of the luminosity during most of the evolution. We examined
the models of \citet{dro} and found nearly linear relationships
between $\log (t_{tAGB} - t)$ and $\log (L / L_{\odot})$ for AGB
evolution between the luminosity minimum in the AGB clump and the
luminosity maximum before the first thermal pulse. ($t_{tAGB}$ is the
age of the star at the luminosity maximum.)  In addition, the slope of
the relation is nearly identical for stars of very different mass, and
is also fairly insensitive to metallicity. The situation is somewhat
complicated by the AGB clump (see \S \ref{sclump} below) and by
thermal pulses (involving interactions between the He and H fusion
shells), but if we restrict ourselves to AGB stars brighter than the
clump, the logarithm of the cumulative LF should also be linear with
magnitude. Because thermal pulses last a very small portion of the
total AGB evolution (10\% or less), they can be safely neglected here.

For cool stars, magnitudes derived from redder filter bands are most
linearly related to $\log (L / L_{\odot})$. For our RGB and AGB
samples we have selected the best $I$ photometry for each star. We
have chosen from the CFHT dataset where available, and from the CTIO
data in all other cases (except for stars outside the CTIO field for
which no $I$ photometry was available). These two sets have the
least scatter in the respective CMDs and are also very nearly on the
same photometric system.

\begin{figure}
\epsscale{.85}
\plotone{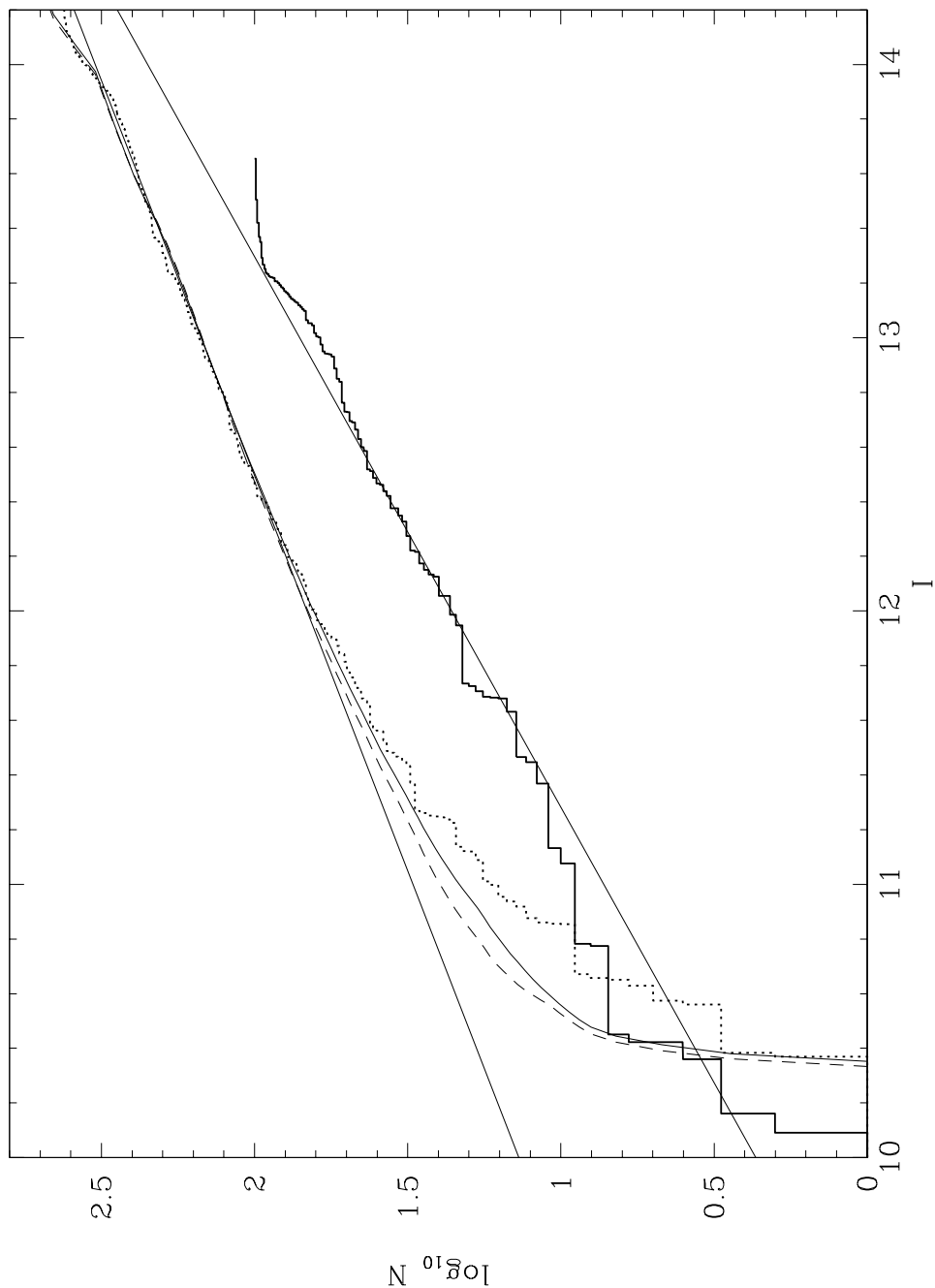}
\caption{Cumulative luminosity functions for AGB {\it thick solid
line} and RGB {\it thick dotted line} stars. The two curves with thin
lines are theoretical predictions from Y$^2$ models \citep{kim} for
[Fe/H] $= -1.11$ and age 12 Gyr with color transformations from Green
et al. (1987; {\it solid line}) and Lejeune et al. (1998; {\it dashed
line}). Also shown are linear fits to the AGB ($I < 12.8$) and RGB
($12 < I < 13.9$). \label{clfi}}
\end{figure}

The cumulative LF in $I$-band for M5 is presented in
Fig. \ref{clfi}. There are two features in this figure that we wish to
discuss: the behavior of the RGB LF at the bright end, and the slopes
of the RGB and AGB LFs.  In the figure we compare with theoretical LFs
for the RGB from Kim et al. (2002; hereafter, Y$^2$ models) for two
different color-$T_{\mbox{eff}}$ transformations. The distance modulus
used in these comparisons was $(m - M)_I = 14.36$ (S96). While the
models reproduce the observed magnitude difference between the RGB tip
and the RGB bump to within 0.1 mag and the slope of the cumulative LF
immediately brighter than the bump ($12 < I < 13.9$), there appear to
be too few stars observed near the red giant tip.  Systematic errors
in the color transformations or underestimation of the neutrino
emission rates used in the stellar models (e.g. Haft, Raffelt, \&
Weiss 1994) might be able to account for at least part of the
difference.  [Changes to neutrino emission rates have other
consequences, however: more than a 50\% change to the rates would
result in greater than a 0.1 mag increase in the brightness of the RGB
tip, which would probably be inconsistent with the observations
(Raffelt \& Weiss 1992).] It should be kept in mind that
identification of RGB and AGB stars becomes most uncertain near the
tip of the RGB where the two branches are closest together in
color. However, the number of ``missing'' RGB stars appears to be
about 8, which is as much as 50\% of the predicted sample at $I <
10.8$. Small horizontal shifts in magnitude do not affect this
discrepancy noticeably. If we use models with [Fe/H] corresponding to
the \citet{zw} scale, the models which match the magnitude level of
the RGB bump also fit most of the upper RGB well, and the slope of the
RGB in the CMD is a closer match to the observations. However, the fit
is unsatisfactory because the tip of the RGB is too faint by 0.4 mag
(9 observed stars fall above the theoretical prediction), and it
requires a distance modulus $(m - M)_I \approx 14.85$ (well outside
the range determined from subdwarf fitting to the main sequence).


To gauge the significance of this difference, we conducted Monte Carlo
simulations. In each trial we randomly chose 300 stars (the number of
observed stars with $I < 13.9$) from a cumulative probability
distribution derived from the theoretical cumulative LF from Y$^2$
models (with Green et al. 1987 color-$T_{\mbox{eff}}$
transformations). We then measured the largest difference in $\log N$
between the simulated and theoretical values at the $I$ magnitude of
each simulated star. The largest deviations in number $N$ tend to
occur near the faint limit of the sample, but the largest fractional
differences (or differences in $\log N$) occur closest to the tip in the
simulations. Even so, in a run of 100,000 trials, we found 1168 trials
with $\log (N_{theory} / N_{sim}) > 0.5$. (The rms value was 0.199.)
If we look for trials having $\log (N_{theory} / N_{sim}) > 0.3$ at $I
> 10.7$, we found 1849 trials of 100,000 meeting the conditions. As a
result, the low number of RGB tip stars observed in M5 is marginally
significant, with a probability of less than 2\% of being due to
statistical fluctuations. In view of the importance to the
understanding of the physics of the upper RGB, other well-populated
clusters should be examined.

The fitted slopes of the AGB and RGB cumulative LFs are significantly
different: $0.3458 \pm 0.0015$ for the RGB above the bump ($12 < I <
13.9$), and $0.496 \pm 0.015$ for the AGB ($I < 12.8$).  The shorter
evolutionary timescale for AGB stars is responsible for differences in
the total numbers of stars, but the slope relates to the change in the
evolutionary timescale as the brightness changes. The RGB slope
agrees closely with predictions from stellar models
\[ \frac{d \log \Phi_{RGB}}{d M_I} \approx
\frac{d \log (t_{tRGB} - t)}{d \log L}\frac{d \log L}{d M_I} \approx
-0.9 / -2.5 = 0.36, \] where $\Phi_{RGB}$ is the
cumulative luminosity function for RGB stars. The comparison with
predictions for the AGB is somewhat more complicated because the
slope has a slight dependence on star mass, with the slope being
minimum for the bluest HB stars that have an AGB phase. From
\citet{dro} models, the predicted slope is still too high:
\[ d \log \Phi_{AGB} / d M_{I} \approx -1.33 / -2.5 = 0.53 ,\] 
where $\Phi_{AGB}$ is the cumulative luminosity
functions for AGB stars.

\begin{figure}
\epsscale{1}
\plotone{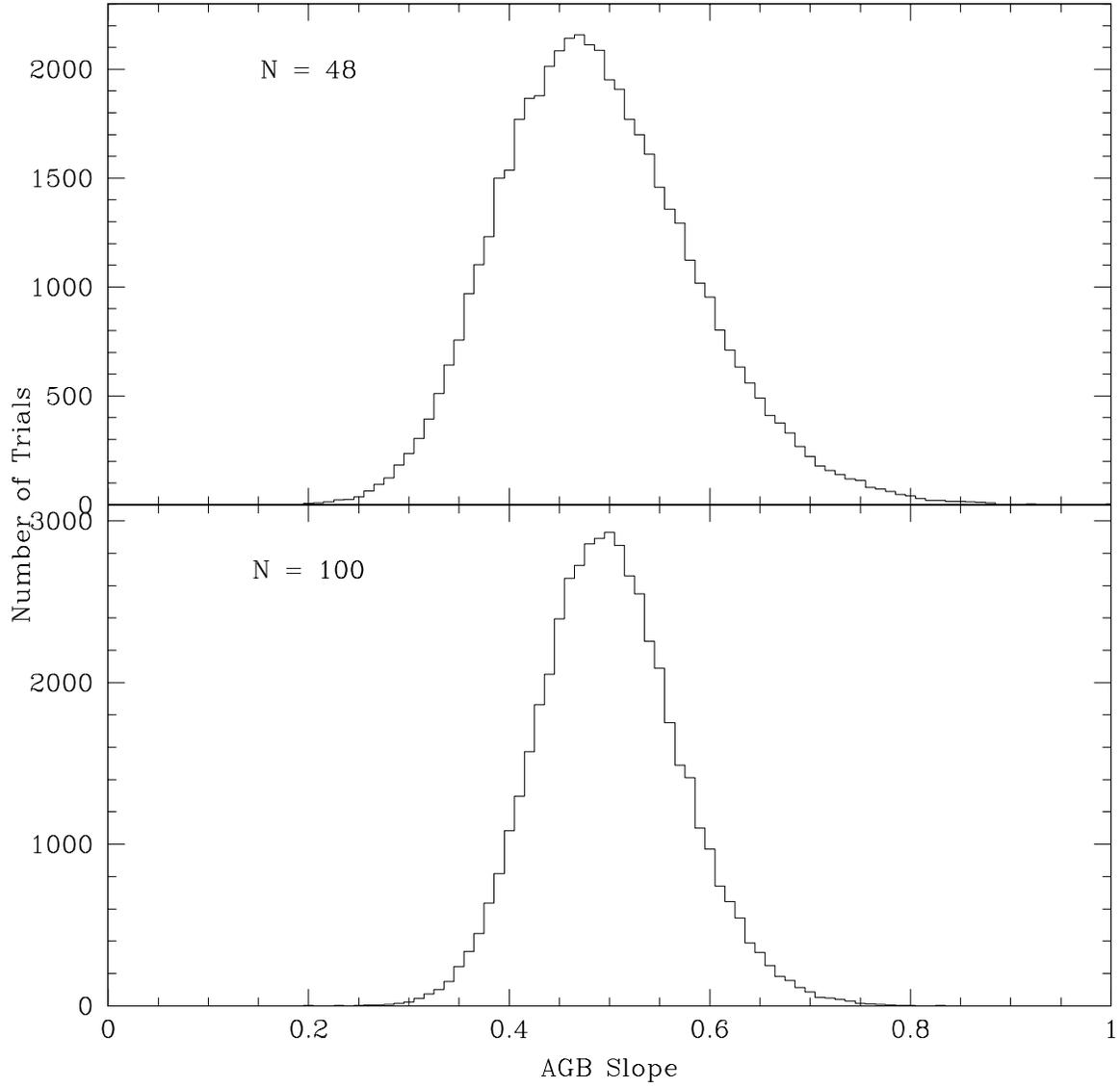}
\caption{Results of Monte Carlo simulations of the slope of the
cumulative luminosity function of bright AGB populations for
48 and 100 stars. \label{agba}}
\end{figure}

To test whether the difference between the observed and predicted
values for the slope of the AGB cumulative LF is significant, we
conducted additional Monte Carlo simulations. We ran 50,000 trials
selecting 48 stars (the number of AGB stars in M5 that were fit to
determine the slope) from a cumulative probability distribution having
a slope $d \log P / dI = 0.53$, and determined an observed slope from
each simulated sample. The trials (see Fig. \ref{agba}) indicate that
there is a significant bias toward measuring a shallower slope (the
average was 0.489, and the mode was 0.47), and the HWHM of the
distribution of slopes was 0.09. The bias comes from stars at the
bright end, which are given somewhat more weight due to the nature of
the cumulative LF: the stars are less abundant, and contribute at a
place in the LF where the addition of a star makes a large fractional
change.

Based on the simulations, our observed slope is quite consistent with
the theoretical prediction, and is in fact almost exactly what is
expected when the bias in measurement is factored in. However, the
simulations indicate that the cumulative LFs are not likely to be able
to give us a strong constraint on AGB evolution in a practical
sense. Fig. \ref{agba} shows the distribution of results for Monte
Carlo trials involving 100 stars on the upper part of the AGB. The
mode of the distribution (0.50) is closer to the input value (0.53),
and the HWHM of the distribution is smaller (0.06). However, there are
only a handful of clusters in the Milky Way that have large enough
populations of bright AGB stars.

\subsection{The AGB Clump}\label{sclump}

With the large AGB sample, we can introduce new
diagnostics to test our understanding of the physics affecting the
evolution of these stars. The AGB phase is characterized by the
progression of a helium fusion shell source through a helium profile
set up during the HB phase. As discussed by \citet{rfp}, this means
than the distribution of stars on the AGB is a reflection of the
helium abundance of the gas being consumed by the shell. The spatial
resolution of the helium fusion shell as a probe of the chemical
profile is poorer than the resolution of the hydrogen fusion shell
during the first ascent of the giant branch because the helium shell
is thicker.  The red giant branch (RGB) bump in globular clusters
provides us with a way of measuring the depth to which the convective
envelope reaches. A similar feature appears at the base of the AGB,
called the ``AGB clump''. During HB evolution, the action of semiconvective
mixing leaves a region of
changing helium abundance that the helium fusion shell processes at
the beginning of the AGB phase. While this happens and the shell
stabilizes with the nearly pure helium mixture outside the mixed
region, an AGB star remains in the clump.

Ferraro et al. (1999) discuss a magnitude difference diagnostic
involving the AGB clump. $\Delta V^{HB}_{AGB} = V^{AGB}_{clump} -
V_{ZAHB}$ is defined in analogy to a more widely used indicator for
the RGB bump: $\Delta V^{bump}_{HB} = V^{RGB}_{bump} -
V_{ZAHB}$. Ferraro et al.  found good agreement with theoretical
models using the FRANEC code \citep{stran} for nine globular clusters,
although this required careful simulation of the HB and determination
of the level of the ZAHB. Because there are no recent sets of HB/AGB
models available, we will not discuss this indicator further here.  In
addition, a earlier comparison of the theoretically predicted
colors of AGB clump stars with observations indicated that the
theoretical predictions were too blue (see Fig. 25 of S96). Again
though, this kind of comparison is somewhat sensitive to composition
and color-$T_{eff}$ transformations, and so we have chosen not to
discuss this here.

The number of stars found in the clump is related to the richness of
the helium in what was the outer convective core. To attempt to test
the evolution of the clump stars, we introduce a new diagnostic $R_{clump}
= N_{clump} / N_{AGB}$, where $N_{clump}$ is the number of stars in
the AGB clump (between the base of the clump and a point 0.25 $V$
magnitudes brighter). Such a definition avoids the need to attempt to
identify the upper boundary of the clump in the AGB samples we
gather. The 0.25 mag size is approximately what is predicted
theoretically for the height of the clump, however.

The base of the AGB clump can be identified using the cumulative LF
presented in \S \ref{clf}. The base of the AGB clump seems to fall at
$V = 14.20$ in the CTIO dataset. There are a small number of stars
fainter than this, but which appear to be better classified as AGB
stars than HB stars. We include these stars in the ``clump'' category,
and find $N_{clump} = 45$, $N_{AGB} = 103$, and $R_{clump} = 0.44 \pm
0.05$.  We have used \[ \sigma^{2}(R_{clump}) = \left(
\frac{R_{clump}}{1+R_{clump}}\right) N_{AGB}^{-1}. \] To make a more
direct connection with theoretical models, we also create a similar
operational definition for $I$-band photometry: stars within 0.30 mag
of the base of the AGB clump. We find that the base of the clump is at
$I = 13.25$ in the M5 data, and using this we get $N_{clump} = 41$,
$N_{AGB} = 99$, and $R_{clump} = 0.41 \pm 0.05$ since the outermost
AGB stars do not have $I$ photometry.

From the models of \citet{dro} shown in Fig. \ref{rvals}, we see
a couple of useful features of this indicator: it is roughly constant
with HB mass for $M \ga 0.56 \msun$, and the values for the two most
metal-rich compositions tabulated are quite similar: $R_{clump}
\approx 0.55$. From this comparison, we find that the observed value
for M5 is a little over $2 \sigma$ too low compared to theory. Once
again, we must remind the reader that the \citet{dro} models have
somewhat dated physical inputs that may result in some systematic
errors.  Until updated models are publicly available, we merely
point out areas where it may be possible to learn new
details about the stellar interiors.

As a final note, the models of \citet{cass} indicate that the
inclusion of breathing pulses in theoretical models decreases the AGB
lifetime by more than 25\%, primarily by considerably shortening the
time spent in the clump. Our results for $R_{2}$ and $R_{clump}$ add
more support for the idea that breathing pulses should somehow be
suppressed in models of HB stars.

\section{Conclusions}

We have compiled a nearly complete list of bright RGB, HB, and AGB
stars for the globular cluster M5 reaching from the core of the
cluster to $8 - 10 \arcmin$ from the center. We have used these
samples to conduct a thorough comparison with theory in order to test
stellar interior physics under the conditions prevalent in these
bright stars. We have introduced a new diagnostic $R_{clump}$ for
evaluating the evolutionary timescale in the early part of the AGB
phase, finding that there is a marginal disagreement between the
observations and the models of \citet{dro}. On the whole we find good
agreement between observations and theory for the cumulative LFs of
the RGB above the RGB bump and for the AGB, indicating that the
evolutionary timescales in both phases are predicted accurately. An
apparent exception appears near the tip of the RGB, where there appear
to be too few giants compared to theoretical predictions. The chance
that this is a statistical fluctuation is less than 2\%. This may be
an indication that the neutrino emission rates in the cores of these
bright stars are underestimated.

The most significant result is the large value for the population
ratio \rat compared to recent theoretical values. The high value is
probably the result of the particular HB morphology of the cluster and
not continuing uncertainties in physical inputs like the cross section
for the \cag reaction or the core mixing algorithm. We encourage new
calculations of HB and AGB phases using updated physics because AGB
stars provide a means of testing predictions for the morphology of
evolutionary tracks for stars fusing He into C and O and of constraining
the \cag reaction rate. In particular,
they stand the greatest chance of identifying the color at which HB
stars change from going into or not going into a typical AGB phase,
and the range of colors for which the AGB phase has its maximum
duration.

We also call attention to a peculiarity of the distribution of stars
on M5's HB. We find that the mass distribution for HB
stars peaks at a position corresponding to the blue edge of the
instability strip. The instability strip is heavily populated due to
the large dispersion in HB masses ($0.02 - 0.03 \msun$). However, the
distribution of stars within the instability strip is heavily biased
toward the red half of the instability strip. This is consistent with
M5 being an Oosterhoff group I cluster, but it means that the first
overtone instability strip is underpopulated compared to the
fundamental strip to the red and compared to the nonvariable stars to
the blue. Because M5 has one of the bluest HB morphologies of the Oo I
clusters (including the more metal-poor M3), it is a severe test of
potential explanations of the Oosterhoff dichotomy. Some mechanism for
reducing the evolutionary timescale of stars in the first overtone
instability strip seems to be needed. The blue half of M5's
instability strip {\it should be} more heavily populated --- the big
question is why it isn't.

\clearpage

\begin{deluxetable}{rrrcccl}
\tablewidth{0pt} 
\tabletypesize{\scriptsize}
\tablecaption{The AGB Star Sample for M5}
\tablehead{\colhead{ID}& \colhead{$\Delta \alpha (\arcsec)$} &
\colhead{$\Delta \delta (\arcsec)$} & \colhead{$P_{\mu}$} & \colhead{Alternate ID}\\ \colhead
{} & \colhead{ID} & & \colhead{$B$}& \colhead{$V$} & \colhead{$I$} &
\colhead{Catalog}}
\startdata
1 & $-135.209$ & $-125.364$ &  0.99 & V42\\
  & 1 & & 12.81 & 10.93 & & PM \\
2 & 7.296 & $-34.384$ & \\
  & 1 & & $13.3947\pm0.0689$ & & $10.0907\pm0.1270$ & CT\\
3 &   6.125 &  86.957 & 0.99 \\
  &    2 & & $13.5448\pm0.0054$&                     & $10.1608\pm0.0730$& CT \\
  &      & & $13.5565\pm0.0050$&  $11.9222\pm0.0025$ & &                   CT \\
  &    2 & & 13.98             &12.02                & &               PM \\
4 &  28.847 & $-72.530$ & \\
  &    5 & & $13.4689\pm0.0142$&                     & $10.3608\pm0.1110$ &CT \\
5 &   7.478 & $-34.628$ & \\
  & 6146 & & $13.6798\pm0.0040$&                     & $10.4222\pm0.0015$ &CFH \\
6 &  35.655 & $-42.208$  & \\
  & 6665 & & $11.7171\pm0.0189$&                     & $10.4512\pm0.0094$ &CFH \\
7 & $-79.898$ & $-28.895$ & 0.99 \\ 
  &   14 & & $13.6607\pm0.0086$&                     & $10.7752\pm0.0740$& CT \\
  &      & & $13.6925\pm0.0090$&  $12.3870\pm0.0024$ & &                   CT \\
  &   10 & &  13.87           &  12.37              & &                 PM \\
8 &  30.902 & $-76.439$ & \\
  &   15 & & $13.6805\pm0.0160$&                     & $10.7819\pm0.0600$&  CT \\
  &      & & $13.6789\pm0.0160$&  $12.2595\pm0.0022$ & &                   CT \\
9 &  53.961 &  69.500 & 0.99 \\
  &32456 & & $13.7817\pm0.0224$&  $12.4399\pm0.0643$ & &                   HST \\
  &  603 & & $13.7703\pm0.0008$&                     & $11.0759\pm0.0035$ & CFH \\
  &   22 & & $13.6924\pm0.0079$&                     & $11.0324\pm0.0310$ & CT \\
  &      & & $13.7025\pm0.0080$&  $12.4456\pm0.0021$& &                    CT   \\
  &   15 & & 13.90             &12.48               & &                PM \\
10 & 114.220 &  61.938 & 0.99 & I-20\\
  &   28 & & $13.8090\pm0.0058$ &                   & $11.1332\pm0.0280$ & CT \\
  &      & & $13.8064\pm0.0060$&  $12.4973\pm0.0014$& &                    CT \\
  &   20 & & 14.02           &  12.62              & &                 PM  \\
\enddata
\label{agb}
\tablecomments{The complete version of this table is in the
electronic edition of the Journal. The printed edition contains only a sample.}
\end{deluxetable}

\begin{deluxetable}{rrrcccl}
\tablewidth{0pt} 
\tabletypesize{\scriptsize}
\tablecaption{The RGB Star Sample for M5}
\tablehead{\colhead{ID}& \colhead{$\Delta \alpha (\arcsec)$} &
\colhead{$\Delta \delta (\arcsec)$} & \colhead{$P_{\mu}$} & \colhead{Alternate ID}\\ \colhead
{} & \colhead{ID} & & \colhead{$B$}& \colhead{$V$} & \colhead{$I$} &
\colhead{Catalog}}
\startdata
 1 & $-32.782$ & $-17.291$ \\ 
  & 4873 &   &$13.7470 \pm  0.0000$ &  &  $10.3694 \pm 0.0069$ & CFH \\
  &    4  &  &$13.7046 \pm  0.0190$      &     &  $10.3198 \pm  0.0530$ & CT \\
  &       &  &$13.7092 \pm  0.0190$ & $12.0561 \pm  0.0027$ &    &  CT\\
 2 &  33.925 &  99.017\\ 
 & 32361  &  &$13.9193 \pm 0.0242$ & $12.3500 \pm 0.0217$ &  &         HST\\
   &   6  &  &$14.0165 \pm 0.0044$   &     &  $10.3842 \pm  0.0980$ & CT\\
    &     &  &$14.0990 \pm 0.0049$ & $12.5138 \pm  0.0019$ &   &       CT\\
 3 &  15.390 & $-44.225$ \\  
 &  6680 &   &$13.7474 \pm 0.0000$ &   &  $ 10.5605 \pm  0.0044$& CFH\\
 &     3 &   &$13.4374 \pm 0.0609$ &   &  $10.2996 \pm 0.1180$ & CT \\
 4 &  $-5.435$  &  0.678 \\
& 210174 &  &$13.7950 \pm 0.0210$ & $12.3154 \pm 0.0395$ &   &   HST\\
  & 3869  &  &$13.7238 \pm  0.0000$ &   &  $10.5741\pm  0.0036$ & CFH\\
 5 & 442.242 & 184.036 & 0.99 \\
   &   3  &  &13.86  &     12.16  &    &              PM \\
 6 & 167.531 & $-175.342$ & 0.99 & IV-81\\
 &     8 &   &$13.7208 \pm 0.0051$ &      &  $10.6289 \pm 0.0430$ &CT\\
    &     &  &$13.7188 \pm 0.0050$ &  $12.1991 \pm 0.0020$  &    &  CT\\
     & 4  &  &13.89    &        12.20 &    &     PM  \\
 7 & $-73.669$ &  45.865 & 0.99\\
 & 23898  &  &$13.8489 \pm 0.0257$ & $12.4448 \pm 0.0313$ &  &  HST\\
 &     9 &   &$13.7404 \pm 0.0076$ &  &  $10.6517 \pm 0.0720$ & CT\\
 &        &  &$13.7438 \pm 0.0080$ & $12.2414 \pm 0.0029$ &   &   CT\\
 &    11  &  &13.93   &   12.38 &     &              PM\\
 8 & $-131.912$ & $-108.340$ & 0.99 & III-122\\
 &    10  &  &$13.7623 \pm 0.0038$ &  & $10.6567 \pm 0.0710$ & CT\\
 &        &  &$13.7774 \pm 0.0039$ & $12.2999 \pm 0.0017$ &   &          CT\\
  &   5   &  & 13.83          &  12.25    &    &   PM  \\
 9 & $-504.784$ & 271.834 & 0.99\\
  &    6  &  & 13.86 &        12.26 &    &                 PM \\
10 &  54.805 &  93.936 \\ 
   &      12 &  & $13.7773 \pm 0.0159$  &    & $10.6717 \pm 0.0560$ &CT\\
    &        &  & $13.7552 \pm 0.0160$  &$12.1734 \pm 0.0029$ & &    CT\\
\enddata
\label{rgb}
\tablecomments{The complete version of this table is in the
electronic edition of the Journal. The printed edition contains only a sample.}
\end{deluxetable}

\begin{deluxetable}{rrrccccccl}
\tablewidth{0pt} 
\tabletypesize{\scriptsize}
\tablecaption{The RR Lyrae Star Sample for M5}
\tablehead{\colhead{ID}& \colhead{$\Delta \alpha (\arcsec)$} &
\colhead{$\Delta \delta (\arcsec)$} & \colhead{$P_{\mu}$} &
\colhead{$V$} & \colhead{$(B-V)_{static}$} & \colhead{Cat.\tablenotemark{a}} &
\colhead{$V$} & \colhead{$(V-I)_{static}$} & \colhead{Notes}}
\startdata
V1  &   25.226 & 150.907 & 0.99 & & & &          15.107 & 0.493 \\ 
V2  & $-353.326$ & $-28.050$ & 0.99 \\
V3  &  156.521 &  99.011 & 0.99 & & & &          15.063 & 0.541 \\
V4  &  $-17.676$ &  65.207 & & & & &             15.070 & 0.413 & HST 23554\\
V5  &  $-13.998$ &  43.050 & & & & &             15.103 & 0.475 & HST 22389\\
V6  &   18.312 & $-56.349$ & & & & &             15.011 & 0.515 \\
V7  &  $-18.581$ & $-199.737$ & 0.99 & 15.035 & 0.344 & S \\
V8  &  122.626 & $-146.600$ & 0.99 & 15.085 & 0.334 & S & 15.072 & 0.514 \\
V9  &  190.772 &  72.110 & 0.99 \\
V10 &  111.964 & 369.089 \\
\enddata
\label{rr}
\tablecomments{The complete version of this table is in the
electronic edition of the Journal. The printed edition contains only a sample.}
\tablenotetext{a}{S: \citet{storm}, B: \citet{brocato}, C: \citet{caputo}}
\end{deluxetable}

\begin{deluxetable}{rrrcccl}
\tablewidth{0pt} 
\tabletypesize{\scriptsize}
\tablecaption{The Nonvariable HB Star Sample for M5}
\tablehead{\colhead{ID}& \colhead{$\Delta \alpha (\arcsec)$} &
\colhead{$\Delta \delta (\arcsec)$} & \colhead{$P_{\mu}$} & \colhead{Alternate ID}\\ \colhead
{} & \colhead{ID} & & \colhead{$B$}& \colhead{$V$} & \colhead{$I$} &
\colhead{Catalog}}
\startdata
  1 & $-577.912$ &  147.750 & 0.99 & II469\\
  &   206 & &   15.32  &           14.58  &     &                        PM \\
  2 & $-549.584$ &  183.246 &  0.99 & II493 \\
  &   342 & &   15.56          &   15.05     &   &                       PM \\
  3 & $-527.766$ & $-39.857$ &  0.99 & II210 \\
  &   465  & &  15.36     &        15.26               &  &              PM\\
  4 & $-521.620$ & $-55.261$ &  0.99 & II213 \\
  &   274 & &  15.04    &         14.95   &  &                           PM\\
  5 & $-501.106$ &  132.187 & 0.99 & II429 \\
   &  263 & &   15.51      &       14.94        & &                       PM\\
  6 & $-444.629$ & $-79.848$ \\
   & 1749 & & $15.6342 \pm 0.0027$ & & $15.7084 \pm 0.0043$ & CT\\
   & & &  $15.6357 \pm 0.0028$  & $15.6777 \pm 0.0018$   & &          CT\\
  7 & $-434.549$ &  128.281  & 0.99 & II425 \\
  &   759 & & $15.5573 \pm 0.0119$ & & $14.4379 \pm 0.0040$ & CT\\
    & & & $15.5576 \pm 0.0120$ & $15.0796 \pm 0.0018$ & &            CT\\
    & 399 & &   15.62  &           15.12  &                 &            PM\\
  8 & $-427.987$ & $-219.067$ &  0.99 \\
  &   773 & & $15.5511 \pm 0.0120$ & & $14.4503 \pm 0.0066$ & CT \\
   &  &  &  $15.5512 \pm 0.0120$ & $15.0818 \pm 0.0010$  &       &       CT\\
  &   381 & &    15.54      &       15.10           &      &              PM\\
  9 & $-395.862$ &  295.727 & 0.99 & I43 \\
 &   1075 & & $15.2769 \pm 0.0150$ & & $14.9241 \pm 0.0052$ &  CT\\
    & & &  $15.2750 \pm  0.0150$ & $15.1301 \pm 0.0017$ & &       CT\\
   &  431 & &   15.24 &            15.15   & &                            PM\\
 10 & $-391.367$ & $-481.471$ &  0.99 \\
   &  454 & &   15.27     &        15.23    & &                           PM \\
\enddata
\label{hb}
\tablecomments{The complete version of this table is in the
electronic edition of the Journal. The printed edition contains only a sample.}
\end{deluxetable}

\clearpage
\begin{deluxetable}{rl}
\tablewidth{0pt} 
\tablecaption{Notes for Bright Star Samples in M5}
\tablehead{\colhead{ID}& \colhead{}}
\startdata
\multicolumn{2}{c}{AGB Stars} \\
1 & (V42) post-AGB star; PM only\\
12 & HST identification ambiguous\\
14 & HST identification ambiguous\\
22 & HST identification ambiguous, blends with RGB 99\\
23 & HST identification ambiguous\\
34 & HST identification ambiguous\\
46 & RGB 310 nearby\\
60 & RGB star (HST 10218) nearby\\
\enddata
\label{notes}
\tablecomments{The complete version of this table is in the
electronic edition of the Journal. The printed edition contains only a sample.}
\end{deluxetable}
\end{document}